\documentclass[runningheads]{svmult}
\usepackage{rotating}  

\usepackage{makeidx}   
\usepackage{graphicx}  
\usepackage{multicol}  
\usepackage{physprbb}  
\makeindex             

\begin{document}

\setcounter{page}{-3}
\thispagestyle{empty}
\begin{center}
\bfseries\Large%
\par\vspace*{0.3\textheight}\par
Five Lectures On Dissipative Master Equations
\par\vspace*{2\baselineskip}\par\large\normalfont%
Berthold-Georg Englert and Giovanna Morigi 
\par\vfill\par\normalsize%
To be published in ``Coherent Evolution in Noisy Environments'',\\
Lecture Notes in Physics, \texttt{http://link.springer.de/series/lnpp/}\\
\copyright\ Springer Verlag, Berlin-Heidelberg-New York\\
\end{center}
\cleardoublepage
\setcounter{tocdepth}{2}\tableofcontents

%
\title*{Five Lectures On Dissipative Master Equations}%
\toctitle{Five Lectures On Dissipative Master Equations}%
\titlerunning{Dissipative Master Equations}%
\author{Berthold-Georg Englert\inst{1,2}
\and Giovanna Morigi\inst{1,3}}%
\authorrunning{Berthold-Georg Englert and Giovanna Morigi}%
\institute{%
Max-Planck-Institut f\"ur Quantenoptik,\\
Hans-Kopfermann-Stra\ss{}e 1,
85748 Garching,
Germany
\and
Department of Mathematics and Department of Physics,\\
Texas A\&M University,
College Station, TX 77843-4242,
U. S. A.
\and
Abteilung Quantenphysik,
Universit\"at Ulm, \\
Albert-Einstein-Allee 11,
89081 Ulm, Germany
}

\maketitle              




\newcommand{\Exp}[1]{\,\mathrm{e}^{\mbox{\footnotesize$#1$}}}
\newcommand{\power}[1]{^{\mbox{\footnotesize$#1$}}}

\newcommand{\DMEtr}[2][]{\,\mathrm{Tr}_{#1}\left\{#2\right\}}

\newcommand{\adj}{^{\dagger}}
\newcommand{\DMEone}{\mathbf{1}}
\newcommand{\cL}{\mathcal{L}}
\newcommand{\cK}{\mathcal{K}}
\newcommand{\cA}{\mathcal{A}}
\newcommand{\cB}{\mathcal{B}}
\newcommand{\cC}{\mathcal{C}}
\newcommand{\cM}{\mathcal{M}}
\newcommand{\cW}{\mathcal{W}}
\newcommand{\dhalf}{{\displaystyle\frac{1}{2}}}
\newcommand{\thalf}{{\textstyle\frac{1}{2}}}

\newcommand{\UP}{\raisebox{-1.8pt}{\parbox[b]{10pt}{%
\begin{picture}(10,8)(0,-1.8)%
\put(0,-1.8){\put(1,2){\line(1,0){8}}%
\put(1,6){\line(1,0){8}}%
\put(5,6){\circle*{2.5}}}%
\end{picture}}%
}}
\newcommand{\DN}{\raisebox{-1.8pt}{\parbox[b]{10pt}{%
\begin{picture}(10,8)(0,-1.8)%
\put(0,-1.8){\put(1,2){\line(1,0){8}}%
\put(1,6){\line(1,0){8}}%
\put(5,2){\circle*{2.5}}}%
\end{picture}}%
}}

\newcommand{\up}{\setlength{\unitlength}{0.8\unitlength}%
\raisebox{-1.8pt}{\parbox[b]{10\unitlength}{%
\begin{picture}(10,8)(0,-1.8)%
\put(0,-1.8){\put(1,2){\line(1,0){8}}%
\put(1,6){\line(1,0){8}}%
\put(5,6){\circle*{2}}}%
\end{picture}}%
}\setlength{\unitlength}{1.25\unitlength}%
}
\newcommand{\dn}{\setlength{\unitlength}{0.8\unitlength}%
\raisebox{-1.8pt}{\parbox[b]{10\unitlength}{%
\begin{picture}(10,8)(0,-1.8)%
\put(0,-1.8){\put(1,2){\line(1,0){8}}%
\put(1,6){\line(1,0){8}}%
\put(5,2){\circle*{2.5}}}%
\end{picture}}%
}\setlength{\unitlength}{1.25\unitlength}%
}

\newcommand{\srup}{\setlength{\unitlength}{0.5\unitlength}%
\raisebox{-1.8pt}{\parbox[b]{10\unitlength}{%
\begin{picture}(10,8)(0,-1.8)%
\put(0,-1.8){\put(1,2){\line(1,0){8}}%
\put(1,6){\line(1,0){8}}%
\put(5,5.3){\makebox(0,0){{\Large$\cdot$}}}}%
\end{picture}}%
}\setlength{\unitlength}{2\unitlength}%
}
\newcommand{\srdn}{\setlength{\unitlength}{0.5\unitlength}%
\raisebox{-1.8pt}{\parbox[b]{10\unitlength}{%
\begin{picture}(10,8)(0,-1.8)%
\put(0,-1.8){\put(1,2){\line(1,0){8}}%
\put(1,6){\line(1,0){8}}%
\put(5,1.3){\makebox(0,0){{\Large$\cdot$}}}}%
\end{picture}}%
}\setlength{\unitlength}{2\unitlength}%
}

\newcommand{\repr}{\mathrel{\widehat{=}}}
\newcommand{\DMEmatr}[5][0pt]{\left(\begin{array}{cc}%
#2 & #3 \\[#1] #4 & #5 \end{array}\right)}
\newcommand{\DMEcol}[3][0pt]{\left(\begin{array}{c}%
#2 \\[#1] #3  \end{array}\right)}
\newcommand{\DMErow}[2]{\left(#1,#2\right)}

\newcommand{\DMEexpect}[1]{\bigl\langle #1 \bigr\rangle}

\newcommand{\DMEeq}[1]{(\ref{DMEeq:#1})}

\newcounter{DMEhw}

\newcommand{\PRA}[3]{Phys.\ Rev.\ A \textbf{#1}, #2 (#3)}
\newcommand{\JMO}[3]{J. Mod.\ Opt.\ \textbf{#1}, #2 (#3)}
\newcommand{\OC}[3]{Opt.\ Commun.\ \textbf{#1}, #2 (#3)}

\newcommand{\DMEind}[1]{#1\index{#1}}
\newcommand{\DMEbegIND}[1]{\index{#1|(}}
\newcommand{\DMEendIND}[1]{\index{#1|)}}
\newcommand{\DMEseealso}[2]{(\emph{see also} #1)\idxquad #2}
\newcommand{\DMEmain}{$\sim$}
\newcommand{\DMEsub}{$\sim\sim$}
\newcommand{\DMExIND}[2]{\index{#1!#2}\index{#2!#1}}



\section*{Introductory Remarks}
\addcontentsline{toc}{section}{\protect\numberline{}Introductory Remarks}
The damped harmonic oscillator is arguably the simplest open quantum system
worth studying.
It is also of great practical importance because it is an essential ingredient
in the theoretical description of many quantum-optical experiments.
One can assume rather safely that the quantum master equation of the simple
harmonic oscillator wouldn't be studied so extensively if it didn't play such
a central role in the quantum theory of lasers and the masers.
Not surprisingly, then, all major textbook accounts of theoretical quantum
optics 
\cite{DMEref:1a,DMEref:1b,DMEref:1c,DMEref:1d,DMEref:1e,DMEref:1f,%
DMEref:1g,DMEref:1h,DMEref:1i,DMEref:1j,DMEref:1k,DMEref:1l,DMEref:1m,%
DMEref:1n,DMEref:1o}
contain a fair amount of detail about damped harmonic oscillators.
Fock state representations or phase space functions of some sort are
invariably employed in these treatments.

The algebraic methods on which we'll focus here are quite different.
They should be regarded as a supplement of, not as a replacement for, the
traditional approaches.
As always, every method has its advantages and its drawbacks:
a particular problem can be technically demanding in one approach, 
but quite simple in another.
This is, of course, also true for the algebraic method.
We'll illustrate its technical power by a few typical examples for which the
standard approaches would be quite awkward.

\section{First Lecture: Basics}
\label{DMEsec:1}
The evolution of a simple damped harmonic oscillator is governed by the 
master equation\index{master equation!of a damped harmonic oscillator}
\begin{eqnarray}
  \label{DMEeq:0}
\frac{\partial}{\partial t}\varrho_t
=\I\omega[\varrho_t,a^{\dagger}a]
&-&\frac{1}{2}A(\nu+1)\left(a^{\dagger}a\varrho_t
-2a\varrho_t a^{\dagger}
+\varrho_t a^{\dagger}a\right)\nonumber\\
&-&\frac{1}{2}A\nu\left(aa^{\dagger}\varrho_t
-2a^{\dagger}\varrho_t a
+\varrho_t aa^{\dagger}\right)\;,  
\end{eqnarray}
where $a\adj,a$ are the ladder operators of the oscillator; $\omega$ is its
natural (circular) frequency; $A$ is the energy decay rate; and $\nu$ is the
number of thermal excitations in the steady state that the statistical operator
$\varrho_t\equiv\varrho_t\bigl(a\adj,a\bigr)$ approaches for very late 
times $t$. 
We'll have much to say about the properties of the solutions of \DMEeq{0},
but first we'd like to give a physical derivation of this equation.

\subsection{Physical Derivation of the Master Equation}\label{DMEsec:1a}
\begin{figure}[!t]
\centering\includegraphics{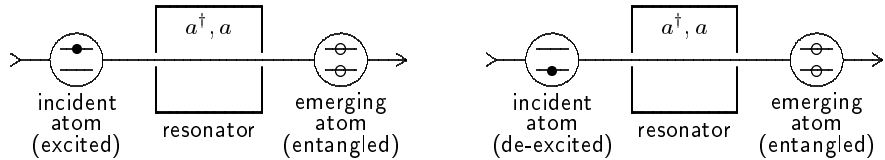}
\caption[A two-level traverses a high-quality cavity \dots\ entangled.]
{\label{DMEfig:1}%
A two-level atom traverses a high-quality cavity, coupling resonantly to a
privileged photon mode of the cavity.
Prior to the interaction, there is some initial photon state 
in the resonator and the atom is either in the upper state 
of the pertinent transition (on the left) 
or in the lower state (on the right).
After the interaction, the transition degree of the atom and the photon
degree of the cavity are entangled}
\end{figure}

For this purpose we consider the following model.
The oscillator is a mode of the quantized radiation field of an ideal
resonator,
so that excitations of this mode (\textit{vulgo\/} photons) would stay in the
resonator forever.
In reality they have a finite lifetime, of course, and we describe this fact
by letting the photons interact with atoms that pass through the resonator.
As is depicted in Fig.~\ref{DMEfig:1}, these atoms are also 
of the simplest kind conceivable:
they only have two levels which -- another simplification -- are separated in
energy by $\hbar\omega$, 
the energy per photon in the privileged resonator mode.
Incident atoms in the upper level (symbolically: \UP) will have a chance to
deposit energy into the resonator, while those in the lower level (\DN) will
tend to remove energy.

The evolution of the interacting atom-photon system is governed by 
the Hamilton operator
\begin{equation}
\label{DMEeq:1}
H=\hbar\omega a^{\dagger}a+\hbar\omega\sigma^{\dagger}\sigma 
-\hbar g(\sigma a^{\dagger}+\sigma^{\dagger}a)\;,
\end{equation}
which goes with the name 
``resonant \DMEind{Jaynes--Cummings interaction} 
in the rotat\-ing-wave approximation'' 
in the quantum-optical literature.
It applies as long as the atom is inside the resonator and is replaced by
\begin{equation}
  \label{DMEeq:1a}
H_\mathrm{free}=\hbar\omega a^{\dagger}a+\hbar\omega\sigma^{\dagger}\sigma 
\end{equation}
before and after the period of interaction.
Here $\sigma\adj$ and $\sigma$ are the atomic ladder operators,
\begin{equation}
  \label{DMEeq:1b}
  \sigma\adj=|\UP\rangle\langle\DN|\repr\DMEmatr{0}{1}{0}{0}\;,\qquad
  \sigma=|\DN\rangle\langle\UP|\repr\DMEmatr{0}{0}{1}{0}\;,
\end{equation}
and $g$ is the so-called \DMEind{Rabi frequency}, 
the measure of the interaction strength.
Note that $\sigma\adj\sigma$ and $\sigma\sigma\adj$ project to the upper and
lower atomic states, 
\begin{equation}
  \label{DMEeq:1c}
  \sigma\adj\sigma=|\UP\rangle\langle\UP|\repr\DMEmatr{1}{0}{0}{0}\;,\qquad
  \sigma\sigma\adj=|\DN\rangle\langle\DN|\repr\DMEmatr{0}{0}{0}{1}\;,
\end{equation}
respectively.

The interaction term in \DMEeq{1}\ is a multiple of the coupling operator
\begin{equation}
  \gamma=a\sigma^{\dagger}+a^{\dagger}\sigma
\end{equation}
and $H_{\mathrm{free}}$ is essentially the square of $\gamma$ since
\begin{equation}
  \gamma^2=a\adj a+\sigma\adj\sigma\;.
\end{equation}
So (\ref{DMEeq:1}) and \DMEeq{1a}\ can be rewritten as
\begin{equation}
\label{DMEeq:2}
H=\hbar\omega\gamma^2-\hbar g\gamma\;,\qquad
H_\mathrm{free}=\hbar\omega\gamma^2\;.
\end{equation}
That $H$ commutes with $H_{\mathrm{free}}$ and both are just simple functions
of $\gamma$ will enable us to solve the equations of motion quite explicitly
without much effort. 

\begin{figure}[!t]
\centering\includegraphics{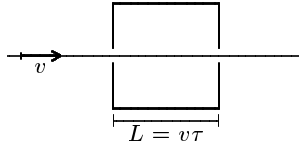}
\caption[An atom takes time $\tau=L/v$ to \dots\ at speed $v$.]
{\label{DMEfig:2}%
An atom takes time $\tau=L/v$ to traverse a cavity of length $L$ at speed $v$}
\end{figure}

We denote by $\varrho_t$ the statistical operator describing the combined 
atom-cavity system.
It is a function of the dynamical variables $a\adj$, $a$, $\sigma\adj$, 
$\sigma$ and has also a parametric dependence on $t$, 
indicated by the subscript,
\begin{equation}
\varrho_t=\varrho_t\bigl(a\adj(t),a(t),\sigma^{\dagger}(t),\sigma(t)\bigr)\;.
\end{equation}
Since a statistical operator has no total time dependence, Heisenberg's
equation of motion,\index{equation of motion!Heisenberg's \DMEmain}
\begin{equation}
  0=\frac{\D}{\D t}\varrho_t
  =\frac{\partial}{\partial t}\varrho_t-\frac{\I}{\hbar}
\bigl[\varrho_t,H\bigr]\;,
\end{equation}
becomes von Neumann's equation for the parametric time dependence,
\index{equation of motion!von Neumann's \DMEmain}
\begin{equation}
\frac{\partial}{\partial t}\varrho_t=\frac\I {\hbar}[\varrho_t,H]\;.
\end{equation}
Now suppose that $t$ is the instant at which the atom enters the cavity; then
it emerges at time $t+\tau$, and we have
\begin{equation}
\label{DMEeq:5}
\varrho_{t+\tau}
=\Exp{-\frac\I {\hbar}H\tau}\varrho_t\Exp{\frac\I {\hbar}H\tau}
=\Exp{-\frac\I {\hbar}H_\mathrm{free}\tau}
\left[\Exp{\I \phi\gamma}\varrho_t\Exp{-\I \phi\gamma}\right]
\Exp{\frac\I {\hbar}H_\mathrm{free}\tau}
\end{equation}
after we use (\ref{DMEeq:2}) and introduce the abbreviation $\phi=g\tau$. 
This phase $\phi$ is the accumulated \DMEind{Rabi angle}
and, for atoms moving classically through the 
cavity of length $L$ with constant velocity $v$,
we have $\tau=L/v$ and $\phi=gL/v$; see Fig.~\ref{DMEfig:2}.

Clearly, the $[\dots]$ term in \DMEeq{5}\ accounts for the net effect of the
interaction, that part of the evolution that happens in addition to the free
evolution generated by $H_{\mathrm{free}}$.
We have
\begin{equation}
  \label{DMEeq:6b}
  \Exp{\I\phi\gamma}
=\cos\left(\phi\gamma\right)+\I\sin\left(\phi\gamma\right)
=\cos\left(\phi\sqrt{\gamma^2}\right)
+\I\gamma\frac{\sin\left(\phi\sqrt{\gamma^2}\right)}{\sqrt{\gamma^2}}\;,
\end{equation}
which is just saying that the cosine is an even function and the sine is odd.
Further we note that the identities
\begin{equation}
  \label{DMEeq:6c}
\begin{array}[b]{rcl}
F\left({\gamma^2}\right)
&=&\sigma\adj\sigma F\left({aa\adj}\right)
+\sigma\sigma\adj F\left({a\adj a}\right)
\;,\\[1ex]
\gamma F\left({\gamma^2}\right)
&=&\sigma a\adj F\left({aa\adj}\right)
+ F\left({aa\adj}\right)a\sigma\adj
\end{array}
\end{equation}
hold for all functions $F\bigl({\gamma^2}\bigr)$.
They are immediate implications of familiar relations such as
$af(a\adj a)=f(aa\adj)a$,
$\sigma f(\sigma\adj\sigma)=\sigma f(1)$,
and $\sigma\adj f(\sigma\adj\sigma)=\sigma\adj f(0)$.
We use \DMEeq{6b} and \DMEeq{6c} to arrive at
\begin{eqnarray}
\Exp{\I\phi\gamma}
&=&\sigma^{\dagger}\sigma\cos\left(\phi\sqrt{aa^{\dagger}}\,\right)+
\sigma\sigma^{\dagger}\cos\left(\phi\sqrt{a^{\dagger}a}\,\right)\nonumber\\
&&\mbox{}+\I\sigma a^{\dagger}
\frac{\sin\left(\phi\sqrt{a a^{\dagger}}\,\right)}{\sqrt{a a^{\dagger}}}
+\I\frac{\sin\left(\phi\sqrt{aa^{\dagger}}\,\right)}{\sqrt{aa^{\dagger}}}a
\sigma^{\dagger}
\;. 
\end{eqnarray}
In terms of the $2\times2$ matrix representation for the $\sigma$'s that is
suggested in \DMEeq{1b} and \DMEeq{1c},
this has the compact form
\begin{equation}
  \label{DMEeq:7}
  \Exp{\I\phi\gamma}\repr\DMEmatr{C}{\I S\adj}{\I S}{\widetilde{C}}
\end{equation}
with the photon operators
\begin{equation}
  C=\cos\left(\phi\sqrt{aa^{\dagger}}\,\right)\;,\qquad
\widetilde{C}=\cos\left(\phi\sqrt{a^{\dagger}a}\,\right)\;,\qquad
S=a\adj
\frac{\sin\left(\phi\sqrt{a a^{\dagger}}\,\right)}{\sqrt{a a^{\dagger}}}\;,
\end{equation}
and the adjoint of \DMEeq{7} reads
\begin{equation}
  \label{DMEeq:7b}
  \Exp{-\I\phi\gamma}\repr\DMEmatr{C}{-\I S\adj}{-\I S}{\widetilde{C}}\;.
\end{equation}

We use these results for calculating the 
net effect of the interaction with one atom on the statistical operator
$\varrho^{(\mathrm{ph})}$ of the photon state.
Initially the total state $\varrho_t=\varrho_t^{\rm(ph)}\varrho_t^{\rm(at)}$ 
is not entangled, it is a product of the statistical operators referring
respectively to the photons by themselves and the atom by itself.
At the final instant $t+\tau$, 
we get $\varrho_{t+\tau}^\mathrm{(ph)}$ by tracing over the two-level atom,
\begin{equation}
\varrho_{t+\tau}^{\rm(ph)}
=\DMEtr[\mathrm{at}]{\varrho_{t+\tau}}
=\Exp{-\I\omega\tau a^{\dagger}a}
\DMEtr[\mathrm{at}]{\Exp{\I\phi\gamma}
\varrho^{(\mathrm{ph})}_t\varrho^{(\mathrm{at})}_t
\Exp{-\I\phi\gamma}}
\Exp{\I\omega\tau a^{\dagger}a}\;.
\label{DMEeq:6}
\end{equation}
To proceed further we need to specify the initial atomic state
$\varrho_t^{(\mathrm{at})}$, and for the present purpose the 
two situations of Fig.~\ref{DMEfig:1} will do.

On the left of Fig.~\ref{DMEfig:1} we have $\UP$ atoms arriving, 
\begin{equation}
\varrho^{(\mathrm{at})}_t
=|\UP\rangle\langle\UP|
=\sigma^{\dagger}\sigma
\repr
\left(\begin{array}{cc} 1 & 0 \\ 0 & 0 \end{array}\right)
=\left(\begin{array}{c} 1 \\ 0 \end{array}\right) (1,0)\;,
\end{equation}
and (\ref{DMEeq:6}) tells us that
\begin{eqnarray}
\varrho_{t+\tau}^{\mathrm{(ph)}}&=&
\Exp{-\I\omega\tau a^{\dagger}a}
\DMEtr[2\times2]{\left(\begin{array}{c} C \\ \I S \end{array}\right)
\varrho_t^{(\mathrm{ph})} (C,-\I S^{\dagger})}
\Exp{\I\omega\tau a^{\dagger}a}\nonumber\\
&=&\Exp{-\I\omega\tau a^{\dagger}a}
\left(C\varrho_t^{(\mathrm{ph})}C+S\varrho_t^{(\mathrm{ph})}S^{\dagger}\right)
\Exp{\I\omega\tau a^{\dagger}a}\;.
\label{DMEeq:A1'}
\end{eqnarray}
Likewise, in the situation on the right-hand side of Fig.~\ref{DMEfig:1} 
we have
\begin{equation}
  \label{DMEeq:A2}
\varrho^{(\mathrm{at})}_t
=|\DN\rangle\langle\DN|
=\sigma\sigma^{\dagger}
\repr\left(\begin{array}{cc} 0 & 0 \\ 0 & 1 \end{array}\right)
=\left(\begin{array}{c} 0 \\ 1 \end{array}\right) (0,1)
\end{equation}
and get
\begin{eqnarray}
\varrho_{t+\tau}^{\mathrm{(ph)}}&=&
\Exp{-\I\omega\tau a^{\dagger}a}
\DMEtr[2\times2]{
\left(\begin{array}{c} \I S^{\dagger}\\ \widetilde{C} \end{array}\right)
\varrho_t^{(\mathrm{ph})} (-\I S,\widetilde{C})}
\Exp{\I\omega\tau a^{\dagger}a}\nonumber\\
&=&\Exp{-\I\omega\tau a^{\dagger}a}
\left(\widetilde{C}\varrho_t^{(\mathrm{ph})}\widetilde{C}
+S^{\dagger}\varrho_t^{(\mathrm{ph})}S\right)
\Exp{\I\omega\tau a^{\dagger}a}\;.
\label{DMEeq:A2'}
\end{eqnarray}

We remember our goal of modeling the coupling of the photons to a reservoir,
and therefore we want to identify the effect of very many atoms traversing the
cavity (one by one) but with each atom coupled very weakly to the photons.
Weak atom-photon interaction means a small value of $\phi$ so that 
only the terms of lowest order in $\phi$ will be relevant.
Since the $\phi=0$ version of both \DMEeq{A1'} and \DMEeq{A2'}, that is:
\begin{equation}
\varrho_{t+\tau}^{\mathrm{(ph)}}=
\Exp{-\I\omega\tau a^{\dagger}a}
\varrho_{t}^{\mathrm{(ph)}}
\Exp{\I\omega\tau a^{\dagger}a}  \;,
\end{equation}
is just the free evolution of $\varrho_{t}^{\mathrm{(ph)}}$, the additional
change in $\varrho_{t}^{\mathrm{(ph)}}$ that results from a single atom is
\begin{eqnarray}
\Delta_1\varrho_{t}^{\mathrm{(ph)}}
&=&C\varrho_{t}^{\mathrm{(ph)}} C+S\varrho_{t}^{\mathrm{(ph)}} S^{\dagger}
-\varrho_{t}^{\mathrm{(ph)}}\nonumber\\
&=&\cos\left(\!\phi\sqrt{aa^{\dagger}}\,\right)\!
\varrho_{t}^{\mathrm{(ph)}}
\cos\left(\!\phi\sqrt{aa^{\dagger}}\,\right)
+a^{\dagger}\frac{\sin\left(\!\phi\sqrt{aa^{\dagger}}\,\right)}
{\sqrt{aa^{\dagger}}} 
\varrho_{t}^{\mathrm{(ph)}} 
\frac{\sin\left(\!\phi\sqrt{aa^{\dagger}}\,\right)}{\sqrt{aa^{\dagger}}}a
\nonumber\\&&\mbox{}-\varrho_{t}^{\mathrm{(ph)}}\nonumber\\
&=&-\frac{1}{2}\phi^2\left(
aa^{\dagger}\varrho_{t}^{\mathrm{(ph)}}
-2 a^{\dagger}\varrho_{t}^{\mathrm{(ph)}} a 
+\varrho_{t}^{\mathrm{(ph)}}aa^{\dagger}
\right) + \mathrm{O}(\phi^4)
\label{DMEeq:B1}
\end{eqnarray}
for a $\UP$ atom arriving, and
\begin{eqnarray}
\Delta_2\varrho_{t}^{\mathrm{(ph)}}
&=&\widetilde{C}\varrho_{t}^{\mathrm{(ph)}}\widetilde{C}
+S^{\dagger}\varrho_{t}^{\mathrm{(ph)}} S
-\varrho_{t}^{\mathrm{(ph)}}\nonumber\\
&=&\cos\left(\!\phi\sqrt{a^{\dagger}a}\,\right)
\varrho_{t}^{\mathrm{(ph)}}
\cos\left(\!\phi\sqrt{a^{\dagger}a}\,\right)
+\frac{\sin\left(\!\phi\sqrt{aa^{\dagger}}\,\right)}{\sqrt{aa^{\dagger}}}a 
\varrho_{t}^{\mathrm{(ph)}} 
a^{\dagger}\frac{\sin\left(\!\phi\sqrt{aa^{\dagger}}\,\right)}
{\sqrt{aa^{\dagger}}}
\nonumber\\&&\mbox{}-\varrho_{t}^{\mathrm{(ph)}}\nonumber\\
&=&-\frac{1}{2}\phi^2\left(
a^{\dagger}a\varrho_{t}^{\mathrm{(ph)}}
-2 a\varrho_{t}^{\mathrm{(ph)}} a\adj 
+\varrho_{t}^{\mathrm{(ph)}}a^{\dagger}a
\right) + \mathrm{O}(\phi^4)
\label{DMEeq:B2}
\end{eqnarray}
for a $\DN$ atom.
So, for weak atom-photon interaction the relevant terms 
in \DMEeq{B1}\ and \DMEeq{B2}\ are the ones proportional to $\phi^2$.

Atoms arriving at statistically independent times (Poissonian 
statistics for the arrival times)
\index{Poissonian statistics!for arrival times}
will thus induce a rate of change of 
$\varrho_{t}^{\mathrm{(ph)}}$ that is given by
\begin{eqnarray}
\frac{\partial}{\partial t}\varrho_t^{(\mathrm{ph})}\Big|_\mathrm{weak}
&=&r_1\Delta_1\varrho_t^{(\mathrm{ph})}+r_2\Delta_2\varrho_t^{(\mathrm{ph})}
\nonumber\\
&=&\mbox{}-\frac{1}{2}r_1\phi^2\left(aa^{\dagger}\varrho_t^{(\mathrm{ph})} 
- 2 a^{\dagger}\varrho_t^{(\mathrm{ph})} a
+ \varrho_t^{(\mathrm{ph})} aa^{\dagger}
\right)\nonumber\\
&&\mbox{}-\frac{1}{2}r_2\phi^2\left(
a^{\dagger}a\varrho_t^{(\mathrm{ph})}
- 2 a\varrho_t^{(\mathrm{ph})} a^{\dagger}
 + \varrho_t^{(\mathrm{ph})} a^{\dagger}a\right)
\end{eqnarray} 
where $r_1$ and $r_2$ are the arrival rates 
for the $\UP$ atoms and the $\DN$ atoms,
respectively.
This is to say that during a period of duration $T$ there will arrive on
average $r_1T$ atoms in state $\UP$ and $r_2T$ atoms in state $\DN$. 

Since the weak interaction with many atoms is supposed to simulate the
coupling to a thermal bath
(temperature $\Theta$),
these rates must be related to each other by a 
\DMEind{Maxwell--Boltzmann factor}, 
\begin{equation}
\label{DMEeq:8}
\frac{r_1}{r_2}=\exp\left(-\frac{\hbar\omega}{k_\mathrm{B}\Theta}\right)
=\frac{\nu}{\nu+1}
\end{equation}
where $\nu>0$ is a convenient parameterization of the temperature.
Also for matters of convenience, we introduce a rate parameter $A$ by writing
$r_1\phi^2=A\nu$, $r_2\phi^2=A(\nu+1)$ and arrive at
\begin{eqnarray}
\frac{\partial}{\partial t}\varrho_t
&=&\frac{\partial}{\partial t}\varrho_t\Big|_\mathrm{free}
+\frac{\partial}{\partial t}\varrho_t\Big|_\mathrm{weak}
\nonumber\\[1ex]
&=&\I\omega[\varrho_t,a^{\dagger}a]
-\frac{1}{2}A(\nu+1)\left(a^{\dagger}a\varrho_t
-2a\varrho_t a^{\dagger}
+\varrho_t a^{\dagger}a\right)\nonumber\\
&&\hphantom{\I\omega[\varrho_t,a^{\dagger}a]}
-\frac{1}{2}A\nu\left(aa^{\dagger}\varrho_t
-2a^{\dagger}\varrho_t a
+\varrho_t aa^{\dagger}\right)
\nonumber\\
&\equiv&\cL\varrho_t\;,
\label{DMEeq:10}
\end{eqnarray}
where the replacement $\varrho_t^{\mathrm{(ph)}}\to\varrho_t$
is made to simplify the notation from here on.
It should be clear that the $\mathrm{O}(\phi^4)$ of \DMEeq{B1} and \DMEeq{B2}
terms are really negligible
in the limiting situation of $r_1,r_2\gg A$ and $\phi^2\ll1$ with 
finite values for the products $r_1\phi^2$ and $r_2\phi^2$. 

Equation \DMEeq{10} is, of course, the master equation \DMEeq{0} that we had
wished to derive by some physical arguments or, at least, make plausible.
From now on, we'll accept it as a candidate for describing a simple damped
harmonic oscillator and study its implications.
These implications as a whole are the ultimate justification for our conviction
that very many crucial properties of damped oscillators are very well modeled
by \DMEeq{10}. 

Before turning to these implications, however, 
we should not fail to mention the obvious. 
The \DMEind{Liouville operator} $\cL$ of \DMEeq{10} is a linear operator: 
the identities\index{Liouville operator!linearity}
\begin{equation}
\cL(\lambda\varrho)=\lambda\cL\varrho\;,\quad 
 \cL(\varrho_1+\varrho_2)=\cL\varrho_1+\cL\varrho_2
\end{equation}
hold for all operators $\varrho$, $\varrho_1$, and $\varrho_2$ 
and all numbers $\lambda$.

\subsection{Some Simple Implications}\label{DMEsec:1b}
As a basic check of consistency let us first make sure 
that \DMEeq{10}\ is not in
conflict with the normalization of $\varrho_t$ to unit total probability, 
that is: $\DMEtr{\varrho_t}=1$ for all~$t$.
Indeed, remembering the cyclic property of the trace, one easily verifies that 
\begin{equation}
  \label{DMEeq:C1}
  \frac{\D}{\D t}\DMEtr{\varrho_t}=\DMEtr{\frac{\partial}{\partial t}\varrho_t}
                            =\DMEtr{\cL\varrho_t}=0\;,
\end{equation}
as it should be. 
Much more difficult to answer is the question if \DMEeq{10}\ preserves the
positivity of $\varrho_t$; we'll remark on that at the end of the third lecture
(see Sect.~\ref{DMEsec:3c} on p.~\pageref{DMEsec:3c}).

Next, as a first application, we determine the time dependence of the
expectation values of the ladder operator $a\adj$, $a$ and the number
operator $a\adj a$.  
Again, the cyclic property of the trace is the tool to be used, and we find
\begin{equation}
  \label{DMEeq:C2}
  \begin{array}[b]{rcl}\displaystyle
 \frac{\D}{\D t}\bigl\langle a\adj\bigr\rangle_t&=&\displaystyle
\frac{\D}{\D t}\DMEtr{a\adj\varrho_t}=
\DMEtr{a\adj\frac{\partial}{\partial t}\varrho_t}
=(\I\omega-\thalf A)\bigl\langle a\adj\bigr\rangle_t\;,\\[2ex]
\displaystyle \frac{\D}{\D t}\bigl\langle a \bigr\rangle_t&=&
\displaystyle(-\I\omega-\thalf A)\bigl\langle a \bigr\rangle_t\;,\\[2ex]
\displaystyle \frac{\D}{\D t}\bigl\langle a\adj a\bigr\rangle_t&=&
\displaystyle-A\bigl(\bigl\langle a\adj a\bigr\rangle_t-\nu\bigr)\;,
  \end{array}
\end{equation}
which are solved by
\begin{equation}
  \label{DMEeq:C3}
  \begin{array}[b]{rcl}\displaystyle
\langle a\adj \rangle_t
&=&\displaystyle
\langle a\adj \rangle_0\Exp{-At/2}\Exp{\I \omega t}\;,\\[2ex]
\langle a \rangle_t
&=&\displaystyle
\langle a \rangle_0\Exp{-At/2}\Exp{-\I \omega t}\;,\\[2ex]
\langle a^{\dagger}a \rangle_t
&=&\displaystyle
\nu+\left(\langle a^{\dagger}a \rangle_0-\nu\right)\Exp{-At}\;,
  \end{array}
\end{equation}
respectively.
Their long-time behavior,
\begin{equation}
  \label{DMEeq:C4}
  t\to\infty\,:\quad\langle a\adj \rangle_t\to0\;,
               \quad\langle a \rangle_t\to0\;,
               \quad\langle a^{\dagger}a \rangle_t\to\nu\;,
\end{equation}
seems to indicate that the evolution comes to a halt eventually.

\subsection{Steady State}
If this is indeed the case, then the master equation \DMEeq{10} must have a
steady state $\varrho^{(\mathrm{ss})}$.
\DMEbegIND{master equation!of a damped harmonic oscillator!steady state}
As we see in \DMEeq{C2}, it is impossible for $\omega$ and $A$ to compensate
for each other and, therefore, $\varrho^{(\mathrm{ss})}$ must commute with the
number operator $a\adj a$, and as a consequence is must be a function of
$a\adj a$ and cannot depend on $a\adj$ and $a$ individually.
Upon writing $f(a\adj a)$ for this function, we have 
\begin{eqnarray}
  \label{DMEeq:C5}
0=\frac{\partial}{\partial t}\varrho^{(\mathrm{ss})} 
=\cL f(a\adj a)
&=&-A(\nu+1)\left[a^{\dagger}a f(a^{\dagger}a)
-af(a^{\dagger}a)a^{\dagger}\right]
\nonumber\\
&&\mbox{}-A\nu\left[aa^{\dagger} f(a^{\dagger}a)
-a^{\dagger}f(a^{\dagger}a)a\right]\;,
\end{eqnarray}
and this implies the three-term recurrence relation
\begin{equation}
  \label{DMEeq:C6}
(a^{\dagger}a+1)\left[(\nu+1)f(a^{\dagger}a+1)
-\nu f(a^{\dagger}a)\right]=
a^{\dagger}a\left[(\nu+1)f(a^{\dagger}a)-\nu f(a^{\dagger}a-1)\right]\;,
\end{equation}
which, incidentally, is a statement about \DMEind{detailed balance} 
(see \cite{DMEref:NathLect} for further details).
In this equation, the left-hand side is obtained from the right-hand side by
the replacement $a\adj a\to a\adj a+1$.
Accordingly, the common value of both sides can be determined by evaluating
the expression for any value that $a\adj a$ may have,
that is: for any of its eigenvalues $(a\adj a)'=0,1,2,\dots$~.
We pick $(a\adj a)'=0$ and find that either side of \DMEeq{C6} must vanish.
The resulting two-term recursion,
\begin{equation}
 (\nu+1)f(a^{\dagger}a)=\nu f(a^{\dagger}a-1) 
\end{equation}
is immediately solved by $f(a\adj a)=f(0)[\nu/(\nu+1)]\power{a\adj a}$ and,
after determining the value of $f(0)$ by normalization, we arrive at
\begin{equation}
\label{DMEeq:11}
\varrho^{(\mathrm{ss})}=\frac{1}{\nu+1}\left(\frac{\nu}{\nu+1}\right)
\power{a^{\dagger}a}
\;.
\end{equation}
This steady state is in fact a \DMEind{thermal state}, 
as we see after re-introducing
the temperature $\Theta$ of \DMEeq{8},
\begin{equation}
\varrho^{(\mathrm{ss})}=
\left[1-\exp\left(-\frac{\hbar\omega}{k_{\mathrm{B}}\Theta}\right)\right]
\exp\left(-\frac{\hbar\omega}{k_{\mathrm{B}}\Theta}a\adj a\right)\;.
\end{equation}
Indeed, together with \DMEeq{C4} this tells us that,
as stated at \DMEeq{0}, ``$\nu$ is the
number of thermal excitations in the steady state''.
And the physical significance of $A$ -- it ``is the energy decay rate'' --
is evident in \DMEeq{C2}.
We might add that $\thalf A$ is the decay rate of the oscillator's 
amplitude  $\langle a\rangle$, which is proportional to the
strength of the electromagnetic field in the optical model.

As the derivation shows, the steady state of \DMEeq{10} is unique, 
unless $A=0$. 
Indeed, if $A=0$ but $\omega\neq0$, $\cL\varrho=0$ is solved by all 
$\varrho=f(a\adj a)$ irrespective of the actual form of the function $f$.
We'll take $A>0$ for granted from here on.
\DMEendIND{master equation!of a damped harmonic oscillator!steady state}

\subsection{Action to the Left}
\DMEbegIND{Liouville operator!action to the left}
In \DMEeq{C2} we obtained differential equations for expectation values from
the equation of motion obeyed by the statistical operator, the master equation
of \DMEeq{10}.
This can be done systematically.
We begin with the expectation value of some observable $X$ and its time
derivative, 
\begin{equation}
  \DMEexpect{X}_t=\DMEtr{X\varrho_t}\;,\qquad
 \frac{\D}{\D t}\DMEexpect{X}_t
=\DMEtr{X\frac{\partial}{\partial t}\varrho_t}\;,
\end{equation}
and then use  \DMEeq{10} to establish
\begin{equation}
   \frac{\D}{\D t}\DMEexpect{X}_t=\DMEtr{X\cL\varrho_t}=\DMEexpect{X\cL}_t\;,
\end{equation}
where the last equation defines the meaning of $X\cL$, that is: the action
of $\cL$ to the left.
The cyclic property of the trace is crucial once more in establishing the
explicit expression
\begin{eqnarray}
  \label{DMEeq:L3}
X\cL
=\I\omega[a^{\dagger}a,X]
&-&\frac{1}{2}A(\nu+1)\left(X a^{\dagger}a
-2a\adj X a
+a^{\dagger}a X\right)\nonumber\\
&-&\frac{1}{2}A\nu\left(Xaa^{\dagger}
-2aXa^{\dagger}
+aa^{\dagger}X\right)\;.  
\end{eqnarray}
When applied to $a\adj$, $a$, and $a\adj a$, this reproduces \DMEeq{C2}, of
course.

How about \DMEeq{C1}? It is also contained, namely as the statement
\begin{equation}
\label{DMEeq:L4}
\DMEtr{\DMEone \cL\varrho_t}=0
\quad\mbox{for all $\varrho_t$, or}\quad\DMEone\cL=0\;,
\end{equation}
which is a statement about the identity operator $\DMEone$.
\DMEendIND{Liouville operator!action to the left}

\subsection*{Homework Assignments}
\addcontentsline{toc}{subsection}{\protect\numberline{}Homework Assignments}
\begin{enumerate}
\renewcommand{\labelenumi}{\textbf{\theenumi}}
\item  
Take the explicit forms of $\Exp{\I\phi\gamma}$ and $\Exp{-\I\phi\gamma}$
in \DMEeq{7}--\DMEeq{7b} and verify that
\begin{equation}  
 \Exp{\I\phi\gamma}\Exp{-\I\phi\gamma}=1\;,\qquad
\Exp{-\I\phi\gamma} \Exp{\I\phi\gamma}=1 \;.
\end{equation}
\item  
According to \DMEeq{C5}, the steady state \DMEeq{11} is a right eigenvector of
$\cL$ with eigenvalue $0$, $\cL\varrho^{(\mathrm{ss})}=0$.
What is the corresponding left eigenvector $\check{\varrho}^{(\mathrm{ss})}$ 
such that $\check{\varrho}^{(\mathrm{ss})}\cL=0$?
\item  
Reconsider \DMEeq{C2} and \DMEeq{C3}.
Show that these equations identify some other eigenvalues of $\cL$ and their
left eigenvectors. 
\item  \label{DMEhw:1d} 
Use the ansatz
\begin{equation}
  \label{DMEeq:2A2}
\varrho_t=\lambda(t)\bigl[1-\lambda(t)\bigr]\power{a^{\dagger}a}
\end{equation}
in the master equation \DMEeq{10},
where $\lambda(t)\to \lambda=1/(1+\nu)$ for $t\to\infty$. 
Derive a differential equation for the numerical function $\lambda(t)$,
and solve it for arbitrary $\lambda(0)$.
[If necessary, impose restrictions on $\lambda(0)$.]
Then recognize that the solution reveals to you some eigenvalues of $\cL$.
Optional: Identify the corresponding right eigenvectors of $\cL$.
\setcounter{DMEhw}{\value{enumi}}
\end{enumerate}


\section{Second Lecture: Eigenvalues and Eigenvectors of $\cL$}
\label{DMEsec:2}

\subsection{A Simple Case First}
\label{DMEsec:2a}

When taking care of homework assignment \ref{DMEhw:1d},
 the reader used \DMEeq{C5} to establish
\begin{equation}
  \cL\varrho_t=-A\frac{1-(\nu+1)\lambda}{1-\lambda}
  \bigl[\lambda a\adj a-(1-\lambda)\bigr]\varrho_t
\end{equation}
for $\varrho_t$ of \DMEeq{2A2} and found
\begin{equation}
  \frac{\partial}{\partial t}\varrho_t
=-\frac{1}{\lambda(1-\lambda)}\frac{\D\lambda}{\D t}
  \bigl[\lambda a\adj a-(1-\lambda)\bigr]\varrho_t
\end{equation}
by differentiation.
Accordingly, \DMEeq{2A2} solves \DMEeq{10} if $\lambda(t)$ obeys
\begin{equation}
  \label{DMEeq:2A4}
  -\frac{1}{\lambda^2}\frac{\D\lambda}{\D t}=\frac{\D}{\D t}\frac{1}{\lambda}
=-A\left(\frac{1}{\lambda}-(\nu+1)\right)\;,
\end{equation}
which is solved by
\begin{equation}
\label{DMEeq:15}
\lambda(t)=\frac{1}{(\nu+1)-\bigl[(\nu+1)-1/\lambda(0)\bigr]\Exp{-At}}
\end{equation}
where the restriction $\lambda(0)>0$ is sufficient to avoid ill-defined values
of $\lambda(t)$ at later times, and $\lambda(0)\leq1$ ensures a positive
$\varrho_t$ throughout.

With \DMEeq{15} in \DMEeq{2A2} we have
\begin{equation}
\varrho_t=\sum_{n=0}^{\infty}\Exp{-nAt}\varrho_n^{(0)}\;,
\label{DMEeq:16}
\end{equation}
where the $\varrho_n^{(0)}$'s are some functions of $a\adj a$.
In particular, $\varrho_0^{(0)}=\varrho^{(\mathrm{ss})}$ is the steady state
\DMEeq{11} that is reached for $t\to\infty$ when $\lambda(t)\to1/(\nu+1)$. 
Since \DMEeq{16} is a solution of the master equation \DMEeq{10} by
construction, it follows that
\begin{equation}
\sum_{n=0}^{\infty}\Exp{-nAt}(-nA)\varrho_n^{(0)}
=\sum_{n=0}^{\infty}\Exp{-nAt}\cL\varrho_n^{(0)}  
\end{equation}
holds for all $t>0$ and, therefore, $\varrho_n^{(0)}$ is a right eigenvector of
$\cL$
\index{Liouville operator!right eigenvectors}%
\index{Liouville operator!eigenvalues}%
with eigenvalue $-nA$,
\begin{equation}
  \cL\varrho_n^{(0)}=-nA\varrho_n^{(0)}\;.
\end{equation}
As defined in \DMEeq{16} with $\varrho_t$ of \DMEeq{2A2} and $\lambda(t)$ of
\DMEeq{15} on the left-hand side, the $\varrho_n^{(0)}$'s depend on the
particular value for $\lambda(0)$, a dependence of no relevance.
We get rid of it by introducing a more appropriate expansion parameter $x$ in
accordance with
\begin{equation}
  \frac{1}{\lambda(t)}=(\nu+1)(1+x)
\quad\mbox{or}\quad
x=\left[\frac{1}{(\nu+1)\lambda(0)}-1\right]\Exp{-At}\;,
\end{equation}
so that counting powers of $\Exp{-At}$ is done by counting powers of $x$.
Then
\begin{equation}
  \label{DMEeq:17c}
\frac{1}{(\nu+1)(1+x)}
\left(1-\frac{1}{(\nu+1)(1+x)}\right)\power{a^{\dagger}a}
=\sum_{n=0}^{\infty}x^n\varrho_n^{(0)}
\end{equation}
is a generating function 
\index{Liouville operator!right eigenvectors!generating function}%
\index{damping bases!generating function}%
\index{generating function!for the damping bases}%
for the $\varrho_n^{(0)}$'s with the spurious
$\lambda(0)$ dependence removed.

The left-hand side of \DMEeq{17c} can be expanded in powers of $x$ for any
value of $\nu\geq0$, but we'll be content with a look at the $\nu=0$ case and
use a different method in Sect.~\ref{DMEsec:2b} to handle the general
situation.
For $\nu=0$, the power series \DMEeq{17c} is simplicity itself,
\begin{eqnarray}
  \label{DMEeq:17d}
\nu=0\,:\quad \sum_{n=0}^{\infty}x^n\varrho_n^{(0)}
&=&\frac{1}{1+x}\left(\frac{x}{1+x}\right)\power{a^{\dagger}a}
\nonumber\\
&=&x\power{a^{\dagger}a}\sum_{m=0}^{\infty} 
{a^{\dagger}a+m \choose m } (-x)^m
\nonumber\\
&=&\sum_{m=0}^{\infty} (-1)^m
{a^{\dagger}a+m \choose a\adj a }x\power{a^{\dagger}a+m}\;, 
\end{eqnarray}
so that
\begin{equation}
  \label{DMEeq:17e}
\nu=0\,:\quad \varrho_n^{(0)}= 
(-1)\power{n-a^{\dagger}a}{n \choose a^{\dagger}a }\;,\quad
\cL\varrho_n^{(0)}=-nA\varrho_n^{(0)}\,. 
\end{equation}
It is a matter of inspection to verify that
\begin{eqnarray}
\varrho_0^{(0)}&=&\delta_{a^{\dagger}a,0}\;,
\nonumber\\
\varrho_1^{(0)}&=&\delta_{a^{\dagger}a,1}-\delta_{a^{\dagger}a,0}\;,
\nonumber\\
\varrho_2^{(0)}&=&\delta_{a^{\dagger}a,2}-2\delta_{a^{\dagger}a,1}
+\delta_{a^{\dagger}a,0}\;,
\nonumber\\ &\vdots&
\end{eqnarray}
are the $\nu=0$ right eigenvectors of $\cL$ to eigenvalues $0,-A,-2A,\dots$~.

We obtain the corresponding left eigenvectors from 
\index{Liouville operator!left eigenvectors}%
\index{damping bases!generating function}%
\index{generating function!for the damping bases}%
\begin{equation}\label{DMEeq:18a}
\nu=0\,:\quad (1+y)\power{a^{\dagger}a}
=\sum_{m=0}^{\infty}y^m\check{\varrho}_m^{(0)}
\end{equation}
after verifying that this is a generating function indeed.
\index{Liouville operator!left eigenvectors!generating function}
For $\nu=0$, \DMEeq{L3} says
\begin{eqnarray}
  \label{DMEeq:18b}
(1+y)\power{a^{\dagger}a}\cL
&=&-Aa\adj a(1+y)\power{a^{\dagger}a}+Aa\adj(1+y)\power{a^{\dagger}a}a 
\nonumber\\
&=&-Aya\adj a(1+y)\power{a^{\dagger}a-1}
\nonumber\\
&=&-Ay\frac{\partial}{\partial y}(1+y)\power{a^{\dagger}a}
\end{eqnarray}
and the eigenvector equation
\begin{equation}
  \check{\varrho}_m^{(0)}\cL=-mA\check{\varrho}_m^{(0)}
\end{equation}
gives
\begin{equation}
  \label{DMEeq:18d}
\sum_{m=0}^{\infty}y^m\check{\varrho}_m^{(0)}\cL=
\sum_{m=0}^{\infty}y^m(-mA)\check{\varrho}_m^{(0)}=
-Ay\frac{\partial}{\partial y}\sum_{m=0}^{\infty}y^m\check{\varrho}_m^{(0)}\;,
\end{equation}
and now \DMEeq{18b} and \DMEeq{18d} establish \DMEeq{18a}.
So we find
\begin{equation}
  \nu=0\,:\quad\check{\varrho}_m^{(0)}={a\adj a \choose m}\;,
\end{equation}
of which the first few are
\begin{eqnarray}
  \label{DMEeq:18f}
  \check{\varrho}_0^{(0)}&=&1\;,\nonumber\\
  \check{\varrho}_1^{(0)}&=&a\adj a\;,\nonumber\\
  \check{\varrho}_2^{(0)}&=&\dhalf a\adj a(a\adj a-1)\;.
\end{eqnarray}
For $n=0$ and $n=1$ this just repeats what was learned in 
\DMEeq{L4} and \DMEeq{C2} (recall homework assignments 2 and 3).

As dual eigenvector sets, the ${\varrho}_n^{(0)}$'s 
and $\check{\varrho}_m^{(0)}$'s
must be orthogonal if $n\neq m$, which is here a statement about the trace of
their product.
It is simplest to deal with them as sets, and we use the generating functions
to establish
\begin{eqnarray}
  \label{DMEeq:19a}
\sum_{m,n=0}^{\infty}y^m
\DMEtr{\check{\varrho}_m^{(0)}\varrho_n^{(0)}}x^n&=&
\DMEtr{(1+y)\power{a^{\dagger}a}\frac{1}{1+x}
\left(\frac{x}{1+x}\right)\power{a^{\dagger}a}}
\nonumber\\
&=&\frac{1}{1+x}\sum_{N=0}^{\infty}\left((1+y)\frac{x}{1+x}\right)^N
\nonumber\\
&=&\frac{1}{1-xy}=\sum_{n=0}^{\infty}x^ny^n
\nonumber\\
&=&\sum_{m,n=0}^{\infty}y^m\delta_{m,n}x^n\;,  
\end{eqnarray}
from which 
\begin{equation}
\label{DMEeq:19}
\mbox{Tr}\left\{\check{\varrho}_m^{(0)}\varrho_n^{(0)}\right\}=\delta_{m,n}
\end{equation}
follows immediately. 
This states the orthogonality of the $\nu=0$ eigenvectors 
and also reveals the sense in which we have normalized them.

In the third lecture we'll convince ourselves of the completeness of the
eigenvector sets. 
Let us take this later insight for granted. 
Then we can write any given initial statistical operator 
$\varrho_{t=0}=f(a\adj a)$
as a sum of the $\varrho_n^{(0)}$,
\begin{equation}
  \label{DMEeq:19b}
  \varrho_{t=0}=\sum_{n=0}^{\infty}\alpha_n^{(0)}\varrho_n^{(0)}\;,
\end{equation}
and solve the master equation \DMEeq{10} by
\begin{equation}
  \label{DMEeq:19c}
  \varrho_t=\sum_{n=0}^{\infty}\alpha_n^{(0)}\Exp{-nAt}\varrho_n^{(0)}\;.  
\end{equation}
As a consequence of the orthogonality relation \DMEeq{19}, we get the
coefficients $\alpha_n^{(0)}$ as
\begin{equation}
  \alpha_n^{(0)}=\DMEtr{\check{\varrho}_n^{(0)}\varrho_{t=0}}\;.
\end{equation}
For $\nu=0$, in particular, they are
\begin{equation}
  \label{DMEeq:19e}
 \alpha_0^{(0)}=1\;,\quad
 \alpha_1^{(0)}=\DMEexpect{a\adj a}_{t=0}\;,\quad
 \alpha_2^{(0)}=\dhalf\DMEexpect{a\adj a(a\adj a-1)}_{t=0}\;,\dots\;, 
\end{equation}
which tells us that \DMEeq{19b} and \DMEeq{19c} are expansions in moments of
the number operator. 
Put differently, the identity
\begin{equation}
  \label{DMEeq:19f}
  f(a\adj a)=\sum_{n=0}^{\infty}
             \varrho_n^{(0)}\DMEtr{\check{\varrho}_n^{(0)}f(a\adj a)}
\end{equation}
holds for any function $f(a\adj a)$ that has finite moments.
For most of the others, one can exchange the roles of $\varrho_n^{(0)}$ and
$\check{\varrho}_n^{(0)}$,
\begin{equation}
  \label{DMEeq:19g}
  f(a\adj a)=\sum_{n=0}^{\infty}
             \DMEtr{f(a\adj a)\varrho_n^{(0)}}\check{\varrho}_n^{(0)}\;.
\end{equation}
A useful rule of thumb 
is to employ expansion \DMEeq{19f} for functions that have the
basic characteristics of statistical operators 
[the traces in \DMEeq{19f} are finite], and use \DMEeq{19g} if
$f(a\adj a)$ is of the kind that is typical for observables
[such as $a\adj a$ for which the $n=0$ trace in \DMEeq{19f} 
is infinite, for example].

\subsection{The General Case}\label{DMEsec:2b}
\DMEbegIND{Liouville operator!right eigenvectors!generating function}
Let us observe that some of the expressions in Sect.~\ref{DMEsec:2a} are of a
somewhat simpler structure when written in normally ordered form
-- all $a\adj$ operators to the left of all $a$ operators --
as exemplified by
\begin{equation}
  \label{DMEeq:20}
  \begin{array}[b]{rcl}
\lambda(1-\lambda)\power{a\adj a}&=&\lambda\;:\Exp{-\lambda a\adj a}:\,,\\[1ex]
(1+y)\power{a\adj a}&=&\;:\Exp{ya\adj a}:\;,\\[1ex] 
\nu=0\,:\quad
\check{\varrho}_m^{(0)}&=&\displaystyle{a^{\dagger}a \choose m}
=\frac{1}{m!}\;:(a\adj a)^m:\;=\frac{{a\adj}^ma^m}{m!}\,.
  \end{array}
\end{equation}
We regard this as an invitation to generalize the ansatz \DMEeq{2A2} to
\begin{equation}
\label{DMEeq:20a}
\varrho_t=\;:\frac{1}{\kappa(t)}
\Exp{-\bigl[a^{\dagger}-\alpha^*(t)\bigr]
\bigl[a-\alpha(t)\bigr]\big/\kappa(t)}:
\end{equation}
where the switching from $\lambda$ to $\kappa=1/\lambda$ 
is strongly suggested by \DMEeq{2A4}. 
Note that for the following it is not required that 
$\alpha^*(t)$ is the complex conjugate of $\alpha(t)$,
it is more systematic to regard them as independent variables.
We pay due attention to the ordering and obtain
\begin{eqnarray}
\frac{\partial}{\partial t}\varrho_t
&=&-\frac{1}{\kappa}\frac{\D\kappa}{\D t}
\varrho_t
+\frac{1}{\kappa^2}\frac{\D\kappa}{\D t}
(a^{\dagger}-\alpha^{*})\varrho_t(a-\alpha)\nonumber\\
&&\mbox{}+\frac{1}{\kappa}\frac{\D\alpha}{\D t}
(a^{\dagger}-\alpha^{*})\varrho_t
+\frac{1}{\kappa}\frac{\D\alpha^*}{\D t}
\varrho_t(a-\alpha)
\end{eqnarray}
for the parametric time derivative of $\varrho_t$.
The evaluation of $\cL\varrho_t$ is equally straightforward once we note that
$a\adj$'s on the right and $a$'s on the left of $\varrho_t$ are moved to their
natural side with the aid of these rules:
\begin{equation}
\begin{array}[b]{rcl}
\varrho_ta\adj&=&a\adj\varrho_t+[\varrho_t,a^{\dagger}]\displaystyle
=a\adj\varrho_t+\frac{\partial}{\partial a}\varrho_t
=a\adj\varrho_t-\frac{1}{\kappa}(a^{\dagger}-\alpha^{*})\varrho_t\;,
\\[1.5ex]
a\varrho_t&=&\varrho_ta+[a,\varrho_t]\displaystyle
=\varrho_ta+\frac{\partial}{\partial a^{\dagger}}\varrho_t
=\varrho_ta-\frac{1}{\kappa}\varrho_t(a-\alpha)
\end{array}
\end{equation}
of which
\begin{equation}
[\varrho_t,a^{\dagger}a]=-\frac{1}{\kappa}(a^{\dagger}-\alpha^{*})\varrho_ta
+\frac{1}{\kappa}a^{\dagger}\varrho_t(a-\alpha)  
\end{equation}
is an immediate application.
Upon equating the numerical coefficients of both $\varrho_t$ and
$(a^{\dagger}-\alpha^{*})\varrho_t(a-\alpha)$ we then get a single equation for
$\kappa(t)$,
\begin{equation}
\label{DMEeq:20d}
\frac{\D\kappa}{\D t}=-A\bigl[\kappa-(\nu+1)\bigr]\;,
\end{equation}
and the coefficients of $(a^{\dagger}-\alpha^{*})\varrho_t$ and
$\varrho_t(a-\alpha)$ supply equations for $\alpha(t)$ and $\alpha^*(t)$,
\begin{equation}
\label{DMEeq:20e}
\frac{\D\alpha}{\D t}
=\left(-\I\omega-\frac{1}{2}A\right)\alpha\;,\quad
\frac{\D\alpha^*}{\D t}
=\left(\I\omega-\frac{1}{2}A\right)\alpha^*\;.
\end{equation}
We have, in fact, met these differential equations before, namely \DMEeq{20d}
in \DMEeq{2A4} and \DMEeq{20e} in \DMEeq{C2}.
Their solutions
\begin{equation}
  \label{DMEeq:20f}
  \begin{array}[b]{rcl}
\kappa(t)&=&\nu+1-(\nu+1-\kappa_0)\Exp{-At}\;,\\[1ex]
\alpha(t)&=&\alpha_0\Exp{-\I\omega t}\Exp{-At/2}\;,\\[1ex]
\alpha^*(t)&=&\alpha_0^*\Exp{\I\omega t}\Exp{-At/2}\;,
  \end{array}
\end{equation}
where $\kappa_0$, $\alpha_0$, $\alpha^*_0$ are the arbitrary initial values at
$t=0$ [not to be confused with the time-independent coefficients
$\alpha_n^{(0)}$ of \DMEeq{19b}--\DMEeq{19e}],
tell us that the time dependence of
\begin{equation}
  \varrho_t=\Exp{\cL t}\varrho_{t=0}
\end{equation}
contains powers of $\Exp{-At}$ combined with powers of 
$\Exp{\mp\I\omega t-At/2}$.
Therefore, the values
\begin{equation}
  \label{DMEeq:20h}
 \lambda_n^{(k)}= -\I k\omega-\bigl(n+\thalf|k|\bigr)A
\quad\mbox{with $n=0,1,2,\dots$ and $k=0,\pm1,\pm2,\dots$}
\end{equation}
must be among the eigenvalues of $\cL$.
\index{Liouville operator!eigenvalues}
In fact, these are all eigenvalues, and the expansion
of the generating function \DMEeq{20a}
\index{damping bases!generating function}%
\index{generating function!for the damping bases}%
\begin{equation} 
\label{DMEeq:21a}
\varrho_t=\sum_{n=0}^{\infty}\sum_{k=-\infty}^{\infty}
b_n^{(k)}\Exp{ -\I k\omega t-\bigl(n+\thalf|k|\bigr)At} \varrho_n^{(k)}
\end{equation}
yields all right eigenvectors of $\cL$,
\index{Liouville operator!right eigenvectors}
\begin{equation}
  \cL\varrho_n^{(k)}=\left[ -\I k\omega-\bigl(n+\thalf|k|\bigr)A\right]
\varrho_n^{(k)}\;.
\end{equation}
We'll justify the assertion that these are \emph{all} in the third lecture.
Right now we just report the explicit expressions that one obtains for the
eigenvectors $\varrho_n^{(k)}$ and the coefficients $b_n^{(k)}$.
They are \cite{DMEref:B+E93,DMEref:ENS94}
\begin{equation} 
\label{DMEeq:21b}
\varrho_n^{(k)}
=\frac{(-1)^n}{(\nu+1)^{|k|+1}}
{a^{\dagger}}\power{\frac{1}{2}(|k|+k)}\;
:\mathrm{L}_n^{(|k|)}\left(\frac{a^{\dagger}a}{\nu+1}\right)
\Exp{-\frac{a^{\dagger}a}{\nu+1}}:\;a\power{\frac{1}{2}(|k|-k)}  
\end{equation}
and
\begin{equation}
  \label{DMEeq:21c}
b_n^{(k)}=\frac{n!}{(n+|k|)!}\left(\frac{\kappa_0}{\nu+1}-1\right)^n
{\alpha_0}\power{\frac{1}{2}(|k|+k)}
\mathrm{L}_n^{(|k|)}\left(\frac{\alpha_0^*\alpha_0}{\nu+1-\kappa_0}\right)
{\alpha_0^*}\power{\frac{1}{2}(|k|-k)}  \;,
\end{equation}
where the $\mathrm{L}_n^{(|k|)}$'s are Laguerre polynomials.
Note that all memory about the initial values of $\kappa$, $\alpha$, and
$\alpha^*$ is stored in $b_n^{(k)}$, as it should be.
The right eigenvectors $\varrho_n^{(k)}$ of \DMEeq{21b} and the left
eigenvectors $\check{\varrho}_n^{(k)}$  of \DMEeq{22b} constitute the
two so-called \emph{damping bases}\index{damping bases} \cite{DMEref:B+E93}
associated with the Liouville operator $\cL$ of \DMEeq{10}.
\DMEendIND{Liouville operator!right eigenvectors!generating function} 

\subsection*{Homework Assignments}
\addcontentsline{toc}{subsection}{\protect\numberline{}Homework Assignments}
\begin{enumerate}
\renewcommand{\labelenumi}{\textbf{\theenumi}}
\addtocounter{enumi}{\value{DMEhw}}
\item 
Verify that \DMEeq{21b} reproduces \DMEeq{17e} for $k=0$ and $\nu=0$.
\item  \label{DMEhw:2b} 
Show that \DMEeq{21a} with \DMEeq{21b} and \DMEeq{21c} is correct.
What is $b_n^{(k)}$ for $\kappa_0=\nu+1$, $\alpha_0=(\nu+1)\beta$, and 
$\alpha_0^*=(\nu+1)\beta^*$?
[You need some familiarity with Bessel functions, Laguerre polynomials, and
relations between them; consult the Appendix if necessary.]
\item  
Consider\index{Liouville operator!left eigenvectors!generating function}
\index{damping bases!generating function}%
\index{generating function!for the damping bases}%
  \begin{equation}
    \label{DMEeq:22a}
    U=\Exp{\alpha a\adj}\Exp{\alpha^* a}\Exp{\lambda}
     =\Exp{\alpha^* a}\Exp{\alpha a\adj}\Exp{\lambda-\alpha^*\alpha}
     =\Exp{\alpha a\adj+\alpha^* a}\Exp{\lambda-\frac{1}{2}\alpha^*\alpha}
  \end{equation}
with $\alpha^*(t)$, $\alpha(t)$, $\lambda(t)$ such that
$\partial U/\partial t=U\cL$.
Find the differential equations obeyed by $\alpha^*$, $\alpha$,
$\lambda$ and solve them.
\item  
Show that this also establishes the eigenvalues \DMEeq{20h} of $\cL$.
\item  \label{DMEhw:2e} 
Extract the left eigenvectors $\check{\varrho}_n^{(k)}$ of $\cL$.
Normalize them such that\index{Liouville operator!left eigenvectors}
\begin{equation}
\label{DMEeq:22b}
\check{\varrho}_n^{(k)}=\left(\frac{-\nu}{1+\nu}\right)^n
\frac{n!}{(n+|k|)!}{a^{\dagger}}\power{\frac{1}{2}(|k|-k)}\;
:\mathrm{L}_n^{(|k|)}\left(\frac{a^{\dagger}a}{\nu}\right):
\;a\power{\frac{1}{2}(|k|+k)}\;.
\end{equation}
What do you get in the limit $\nu\to0$?
\item  
State $\check{\varrho}_n^{(k)}$ explicitly for $k=0$, $n=0,1,2$ and for
$k=\pm1,\pm2$, $n=0$. 
Compare with \DMEeq{18f}.
\item  
Use the two generating functions \DMEeq{20a} and \DMEeq{22a} to demonstrate 
\begin{equation}
\label{DMEeq:22c}
\DMEtr{\check{\varrho}_m^{(k)}\varrho_{n}^{(k')}}
=\delta_{m,n}\delta_{k,k'}\;.
\end{equation}
\setcounter{DMEhw}{\value{enumi}}
\end{enumerate}


\section{Third Lecture: Completeness of the Damping Bases}

\subsection{Phase Space Functions}
As a preparation we first consider the standard phase space functions 
$f(Q',P')$ associated with an operator $F(Q,P)$
where $Q,P$ are a Heisenberg pair 
(colloquially: position $Q$ and momentum $P$).
This is to say that they obey the commutation relation
\begin{equation}
  [Q,P]=\I
\end{equation}
and have complete, orthonormal sets of eigenkets and eigenbras,
\begin{equation}
  \label{DMEeq:23b}
  \begin{array}[b]{rcl@{\qquad}rcl}
Q|Q'\rangle&=&|Q'\rangle Q'\;, & P|P'\rangle&=&|P'\rangle P'\;,\\[1ex]
\langle Q'|Q&=&Q'\langle Q'|\;, & \langle P'|P&=&P'\langle P'|\;,\\[1ex]
\langle Q'|Q''\rangle&=&\delta(Q'-Q'')\;, &
\langle P'|P''\rangle&=&\delta(P'-P'')\;, \\[1ex]
|Q'\rangle\langle Q'|&=&\delta(Q-Q')\;, &
|P'\rangle\langle P'|&=&\delta(P-P')\;, \\[1ex]
\displaystyle\int\!\D Q'\,|Q'\rangle\langle Q'|
&=&1\;,& 
\displaystyle\int\!\D P'\,|P'\rangle\langle P'|
&=&1\;,
\end{array}
\end{equation}
where $Q',Q''$ and $P',P''$ denote eigenvalues.
The familiar plane waves
\begin{equation}
  \langle Q'|P'\rangle=\frac{\Exp{\I Q'P'}}{\sqrt{2\pi}}
\end{equation}
relate the eigenvector sets to each other.

By using both completeness relations, we can put any $F=F(P,Q)$ into its
$Q,P$-ordered form,\index{ordered operator}
\begin{eqnarray}
F(P,Q)
&=&\displaystyle
\int\!\D Q'\,\D P'\,|Q'\rangle\langle Q'|F|P'\rangle\langle P'|
\nonumber\\
&=&
\displaystyle
\int\!\D Q'\,\D P'\,
|Q'\rangle\langle Q'|
\left[\frac{\langle Q'|F|P'\rangle}{\langle Q'|P'\rangle}\right]
|P'\rangle\langle P'|
\nonumber\\
&=&
\displaystyle
\int\!\D Q'\,\D P'\,
\delta(Q- Q')f(Q',P')\delta(P-P')
\nonumber\\
&=&\displaystyle
f(Q,P)\Big|_{Q,P\mathrm{-ordered}}\equiv f(Q;P)\;,
\label{DMEeq:24a}
\end{eqnarray}
where the last step defines the meaning of the semicolon in $f(Q;P)$. 
Thus, the procedure is this:
evaluate the normalized matrix element
\begin{equation}
f(Q',P')
=\frac{\langle Q'|F|P'\rangle}{\langle Q'|P'\rangle}
=\DMEtr{F\frac{|P'\rangle\langle Q'|}{\langle Q'|P'\rangle}}\;,
\label{DMEeq:24b}
\end{equation}
then replace $Q'\to Q$, $P'\to P$ with due attention to their order in
products -- all $Q$'s must stand to the left of all $P$'s --
and so obtain $F=f(Q;P)$, the $Q,P$-ordered form of $F$.
The $P,Q$-ordered version of $F$ is found by an analogous procedure with the
roles of $Q$ and $P$ interchanged. 

The fraction that appears in the trace formula of \DMEeq{24b} is equal to its
square,
\begin{equation}
  \left[\frac{|P'\rangle\langle Q'|}{\langle Q'|P'\rangle}\right]^2
=\frac{|P'\rangle\langle Q'|}{\langle Q'|P'\rangle}
\end{equation}
but, not being Hermitian, it is not a projector.
It has, however, much in common with projectors, 
and this is emphasized by using
\begin{equation}
  \frac{1}{\langle Q'|P'\rangle}=2\pi\langle P'|Q'\rangle
\end{equation}
to turn it into
\begin{equation}
  \frac{|P'\rangle\langle Q'|}{\langle Q'|P'\rangle}=
2\pi|P'\rangle\langle P'|Q'\rangle\langle Q'|
=2\pi\delta(P-P')\delta(Q-Q')\;,
\end{equation}
which is essentially the product of two projectors.
Then,
\begin{equation}
  \label{DMEeq:24f}
f(Q',P')=\DMEtr{F\,2\pi\delta(P-P')\delta(Q-Q')}  
\end{equation}
is yet another way of presenting $f(Q',P')$.

In \DMEeq{24a} and \DMEeq{24f} we recognize two basis sets of operators,
\index{operator basis}
\begin{eqnarray}
 B(Q',P')&=&2\pi\delta(Q-Q')\delta(P-P') \;,\nonumber\\
\widetilde{B}(Q',P')&=&2\pi\delta(P-P')\delta(Q-Q')\;,
\end{eqnarray}
labeled by the phase space variables $Q'$ and $P'$.
Their dual roles are exhibited by the compositions of  
\DMEeq{24a} and \DMEeq{24f},
\begin{equation}
  F(Q,P)=\int\!\frac{\D Q'\,\D P'}{2\pi}
\,B(Q',P')\DMEtr{F\,\widetilde{B}(Q',P')}
\end{equation}
and
\begin{equation}
  f(Q',P')=\DMEtr{\widetilde{B}(Q',P')
\int\!\frac{\D Q''\,\D P''}{2\pi}\,f(Q'',P'')B(Q'',P'')}\;,
\end{equation}
and therefore
\begin{equation}
\label{DMEeq:25d}
\DMEtr{B(Q',P')\widetilde{B}(Q'',P'')}=2\pi\delta(Q'-Q'')\delta(P'-P'')
\end{equation}
states both their orthogonality and their completeness.

\begin{figure}[!t]
\centering\includegraphics{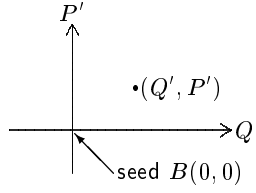}
\caption[The unitary displacements \dots\ of the basis.]{\label{DMEfig:3}%
The unitary displacements $Q\to Q-Q'$, $P\to P-P'$ 
map out all of phase space, turning the basis seed 
$B(0,0)$ into all other operators $B(Q',P')$ of the basis}
\end{figure}

We note that the displacements
\begin{equation}
  Q\to Q-Q'\;,\qquad P\to P-P'
\end{equation}
that map out all of phase space (see Fig.~\ref{DMEfig:3}) 
are unitary operations, and so is the interchange
\begin{equation}
Q\to P\;,\quad P\to -Q\;.  
\end{equation}
As a consequence, $B(Q',P')$ and $\widetilde{B}(Q',P')$ are unitarily
equivalent to their respective basis seeds
\index{operator basis!seed of \DMEmain}
\begin{equation}
  B(0,0)=2\pi\delta(Q)\delta(P)
\quad\mbox{and}\quad
  \widetilde{B}(0,0)=2\pi\delta(P)\delta(Q)\;,
\end{equation}
and these seeds themselves are unitarily equivalent and are also adjoints of
each other,
\begin{equation}
  B(0,0)\adj=\widetilde{B}(0,0)\;,\quad\widetilde{B}(0,0)\adj=B(0,0)\;,
\end{equation}

Here, $B(0,0)$ is stated as a $Q,P$-ordered operator and $\widetilde{B}(0,0)$
is $P,Q$-ordered.
The reverse orderings are also available, as we illustrate for $B(0,0)$:
\begin{equation}
 \frac{\langle P'|B(0,0)|Q'\rangle}{\langle P'|Q'\rangle}
=2\pi \frac{\langle P'|Q''\rangle\langle Q''|P''\rangle
\langle P''|Q'\rangle}{\langle P'|Q'\rangle}\bigg|_{Q''=0,\ P''=0}
=\Exp{\I P'Q'}\;,
\end{equation}
giving
\begin{equation}
  B(0,0)=\Exp{\I P;Q}=\sum_{k=0}^{\infty}\frac{\I^k}{k!}P^kQ^k\;,
\end{equation}
where we meet a typical \DMEind{ordered exponential operator} function.
We get the $Q,P$-ordered version of  $\widetilde{B}(0,0)$,
\begin{equation}
   \widetilde{B}(0,0)=\Exp{-\I Q;P}\;,
\end{equation}
by taking the adjoint.

These observations can be generalized in a simple and straightforward manner
\cite{DMEref:BGE89}.
The operators
\begin{eqnarray}
\label{DMEeq:27a}
&\lambda\Exp{\I \lambda P;Q}
=\bar{\lambda}\Exp{-\I \bar{\lambda} Q;P}\;,\quad
\lambda\Exp{-\I \lambda Q;P}
=\bar{\lambda}\Exp{\I \bar{\lambda}P;Q}&\nonumber\\
&\mbox{with $\lambda,\bar{\lambda}\ge 1$ 
and $\lambda\bar{\lambda}=\lambda+\bar{\lambda}$}&
\end{eqnarray}
are seeds of a dual pair of bases
\index{operator basis!dual pair}
 for each choice of 
$\lambda,\bar{\lambda}$ because the orthogonality-completeness relation
\begin{equation}
\DMEtr{\lambda\Exp{\I \lambda (P-P');(Q-Q')}
\lambda\Exp{-\I \lambda (Q-Q'');(P-P'')}}=
2\pi\delta(Q'-Q'')\delta(P'-P'')
\end{equation}
holds generally, not just for $\lambda=1,\bar{\lambda}=\infty$ as we've seen
in \DMEeq{25d}.
We demonstrate the case by using the two completeness relations of \DMEeq{23b}
to turn the operators into numbers, and then recognize two Fourier
representation of Dirac's $\delta$ function:
\begin{eqnarray}
\DMEtr{\vphantom{\Exp{Q}}\cdots}&=&
\int\frac{\D\bar{Q}\,\D\bar{P}}{2\pi}\lambda
\Exp{\I\lambda(\bar{P}-P')(\bar{Q}-Q')}\lambda
\Exp{-\I\lambda(\bar{Q}-Q'')(\bar{P}-P'')}
\nonumber\\
&=&2\pi\Exp{\I\lambda(Q'P'-Q''P'')}\!
\int\!\frac{\D\bar{P}}{2\pi}\lambda\Exp{-\I\lambda\bar{P}(Q'-Q'')}
\!\int\!\frac{\D\bar{Q}}{2\pi}\lambda\Exp{-\I\lambda\bar{Q}(P'-P'')}
\nonumber\\
&=&2\pi\Exp{\I\lambda(Q'P'-Q''P'')}
\delta(Q'-Q'')\delta(P'-P'')
\nonumber\\
&=&2\pi\delta(Q'-Q'')\delta(P'-P'')\;,
\end{eqnarray}
indeed.
The equivalence of the $\lambda$ and $\bar{\lambda}$ versions of \DMEeq{27a}
is the subject matter of homework assignment~\ref{DMEhw:3a}.

The basis seeds \DMEeq{27a} can be characterized by the similarity
transformations they generate,
\begin{eqnarray}
  \label{DMEeq:27d}
       Q \;\lambda\Exp{\I \lambda P;Q}
    =\lambda\Exp{\I \lambda P;Q}\;(1-\lambda)Q 
&\quad\mbox{or}\quad&Q\to(1-\lambda)Q\;,
\nonumber\\
  P \;\lambda\Exp{\I \lambda P;Q}
   =\lambda\Exp{\I \lambda P;Q}\;(1-\bar{\lambda})P
&\quad\mbox{or}\quad&
P\to(1-\bar{\lambda})P\;,
\end{eqnarray}
which are \DMEind{scaling transformations} essentially (but not quite because
$1-\lambda=1/(1-\bar{\lambda})<0$ and the cases $\lambda=1$ or
$\bar{\lambda}=1$ are particular).
One verifies \DMEeq{27d} with the aid of identities such as
\begin{equation}
   P\,\lambda\Exp{\I \lambda P;Q}
=\frac{1}{\I}\frac{\partial}{\partial Q}\Exp{\I \lambda P;Q}
=\left[P,\Exp{\I \lambda P;Q}\right]\;.
\end{equation}
Equations \DMEeq{27d} by themselves determine the seed only up to an over-all
factor, and this ambiguity is removed by imposing the normalization to unit
trace,
\begin{equation}
  \label{DMEeq:27f}
  \DMEtr{\lambda\Exp{\I \lambda P;Q}}
=\int\frac{\D\bar{Q}\,\D\bar{P}}{2\pi}\lambda
\Exp{\I\lambda \bar{P}\bar{Q}}=1\;.
\end{equation}

The cases $\lambda=1,\bar{\lambda}\to\infty$ and  
$\lambda\to\infty,\bar{\lambda}=1$ are just the bases associated with the 
$Q,P$-ordered and $P,Q$-ordered phase space functions that we discussed above.
Of particular interest is also the symmetric case of $\lambda=\bar{\lambda}=2$
which has the unique property that the two bases are really just one: The
operator basis underlying Wigner's phase space function.
\index{Wigner function}
Its seed (not seed\underline{s}!) is Hermitian, since
$\lambda=\bar{\lambda}=2$ in \DMEeq{27a} implies
\begin{equation}
  \label{DMEeq:28a}
2\Exp{\I2P;Q}=  2\Exp{-\I2Q;P}= \left[2\Exp{\I2P;Q}\right]\adj\;. 
\end{equation}
It then follows, for example, that $F=F\adj$ has a real Wigner function;
in fact, all of the well known properties of the much studied Wigner functions
can be derived rather directly from the properties of this seed (see
\cite{DMEref:BGE89} for details). 

For our immediate purpose we just need to know the following.
When $\lambda=\bar{\lambda}=2$, the transformation \DMEeq{27d} is the
inversion 
\begin{equation}
  \label{DMEeq:28a'}
  Q\to-Q\;,\qquad P\to-P\;.
\end{equation}
For $a\adj=2^{-1/2}(Q-\I P)$, $a=2^{-1/2}(Q+\I P)$ this means that
the number operator $a\adj a=\thalf(Q^2+P^2-1)$ is invariant.
Put differently, the Wigner seed \DMEeq{28a} commutes with $a\adj a$, it is a
function of $a\adj a$: $2\Exp{\I2P;Q}=f(a\adj a)$.
For $a$, the inversion \DMEeq{28a'} requires $af(a\adj a)=-f(a\adj a)a$, 
and this combines with $af(a\adj a)=f(a\adj a+1)a$ to tell us 
that $f(a\adj a+1)=-f(a\adj a)$.
We conclude that
\begin{equation}
    \label{DMEeq:28b}
    2\Exp{\I2P;Q}=2(-1)\power{a\adj a}=\;:2\Exp{-2a\adj a}:
\end{equation}
after using the normalization \DMEeq{27f} to determine the prefactor of $2$,
\begin{equation}
  \DMEtr{2(-1)\power{a\adj a}}
=2\sum_{n=0}^{\infty}(-1)^n
=2\sum_{n=0}^{\infty}(-x)^n\bigg|_{1>x\to 1}
=\frac{2}{1+x}\bigg|_{x\to 1}=1\;.
\end{equation}
The last, normally ordered, version in \DMEeq{28b} of the Wigner seed 
\index{operator basis!seed of Wigner's \DMEmain}
shows the $\lambda=2$ case of \DMEeq{20}.

Perhaps the first to note the intimate connection between the Wigner function
\index{Wigner function!and inversion}
and the inversion \DMEeq{28a'} and thus to recognize that the Wigner seed is
(twice) the parity operator $(-1)\power{a\adj a}$
\index{Wigner function!and parity operator}
\index{parity operator}%
\index{parity operator!and Wigner function}%
was Royer \cite{DMEref:Royer}.
In the equivalent language of the Weyl quantization scheme the analogous
observation was made a bit earlier by Grossmann \cite{DMEref:Grossmann}. 
A systematic study from the viewpoint of operator bases is given in 
\cite{DMEref:BGE89}.

We use the latter form \DMEeq{28b} to write an operator $F(a\adj,a)$ 
in terms of its Wigner function $f(z^*,z)$,
\begin{eqnarray}
 F(a\adj,a)=\int \!\frac{\D Q'\,\D P'}{2\pi}f(z^*,z)
\;:2\Exp{-2\bigl(a\adj-z^*\bigr)\bigl(a-z\bigr)}:\;,
\end{eqnarray}
where $z^*=2^{-1/2}(Q'-\I P')$ and $z=2^{-1/2}(Q'+\I P')$ are
understood, and 
\begin{eqnarray}
  f(z^*,z)
=\DMEtr{F(a\adj,a)\;
:2\Exp{-2\bigl(a\adj-z^*\bigr)\bigl(a-z\bigr)}:}
\end{eqnarray}
reminds us of how we get the phase space function by tracing the products with
the operators of the dual basis (which, we repeat, is identical to the
expansion basis in the $\lambda=\bar{\lambda}=2$ case of the Wigner basis).

\subsection{Completeness of the Eigenvectors of $\cL$}
\DMEbegIND{Liouville operator!completeness of eigenvectors}
\DMEbegIND{damping bases!completeness}
The stage is now set for a demonstration of the completeness of the 
damping bases of Sect.~\ref{DMEsec:2b}.
We'll deal explicitly with the right
eigenvectors $\varrho_n^{(k)}$ of the Liouville operator $\cL$ of \DMEeq{10}
as obtained in \DMEeq{21b} by expanding the generating function \DMEeq{20a}.
Consider some arbitrary initial state $\varrho_{t=0}$ and its Wigner function
representation
\begin{equation}
 \varrho_{t=0}=\int \!\frac{\D Q'\,\D P'}{2\pi}\rho(z^*,z)
\;:2\Exp{-2\bigl(a\adj-z^*\bigr)\bigl(a-z\bigr)}:\;,
\end{equation}
where $\rho(z^*,z)$ is the Wigner phase space function of
$\varrho_{t=0}$.
Then, according to Sect.~\ref{DMEsec:2b}, we have at any later time
\begin{equation}
  \label{DMEeq:29b}
\varrho_t=\Exp{\cL t} \varrho_{t=0}
=\int \!\frac{\D Q'\,\D P'}{2\pi}\rho(z^*,z)
\;:\frac{1}{\kappa(t)}
\Exp{-\bigl[a^{\dagger}-\alpha^*(t)\bigr]
\bigl[a-\alpha(t)\bigr]\big/\kappa(t)}:
\end{equation}
with
\begin{equation}
\kappa_0=\thalf\;,\quad\alpha_0=z\;, 
\quad\alpha_0^*=z^*  
\end{equation}
in \DMEeq{20f}.
In conjunction with \DMEeq{21a}, \DMEeq{21b}, and \DMEeq{21c}
this gives
\begin{equation}
  \label{DMEeq:29d}
\varrho_t=\sum_{n=0}^{\infty}\sum_{k=-\infty}^{\infty}
\beta_n^{(k)}\Exp{ -\I k\omega t-\bigl(n+\thalf|k|\bigr)At} 
\varrho_n^{(k)}  
\end{equation}
with
\begin{eqnarray}
\beta_n^{(k)}&=&\frac{n!}{(n+|k|)!}
\left(-\frac{\nu+\frac{1}{2}}{\nu+1}\right)\power{n}
\nonumber\\[1ex]
&&\times\!\int \!\frac{\D Q'\,\D P'}{2\pi}\rho(z^*,z)  
z\power{\frac{1}{2}(|k|+k)}
\mathrm{L}_n^{(|k|)}\left(\frac{z^*z}{\nu+\frac{1}{2}}\right)
{z^*}\power{\frac{1}{2}(|k|-k)}.
\end{eqnarray}
But this is just to say that any given $\varrho_t$ has an expansion 
in terms of the $\varrho_n^{(k)}$ for all $t$ -- any $\varrho_t$ can be
expanded in the right damping basis.
Equation \DMEeq{29d} also confirms that the exponentials 
$\exp\bigl(-\I k\omega t-\bigl(n+\thalf|k|\bigr)At\big)
=\exp\bigl(\lambda_n^{(k)}t\bigr)$
constitute all possible time dependences that $\varrho_t$ might have.
Clearly, then, the $\lambda_n^{(k)}$'s of \DMEeq{20h} are \emph{all}
eigenvalues of $\cL$, indeed, and 
the right eigenvectors $\varrho_n^{(k)}$ of \DMEeq{21b} are complete.
The completeness of the left eigenvectors $\check{\varrho}_n^{(k)}$ 
of \DMEeq{22b} can be shown similarly, or can be inferred from \DMEeq{22c}.

Note that this argument does not use any of the particular properties that
$\varrho_{t=0}$ might have as a statistical operator.
All that is really required is that its Wigner function
$\rho(z^*,z)$ exists, and this is almost no requirement at all,
because only operators that are singularly pathological may not possess a
Wigner function. 
Such exceptions are of no interest to the physicist.

More critical are those operators $\varrho_{t=0}$ for which the expansion is
of a more formal character because the resulting coefficients $\beta_n^{(k)}$
are distributions, rather than numerical functions, of the parameters that are
implicit in $\rho(z^*,z)$.
In this situation the recommended procedure is to expand in terms of the left
eigenvectors $\check{\varrho}_n^{(k)}$ instead of the right eigenvectors
$\varrho_n^{(k)}$.  

The demonstration of completeness given here relies on machinery developed in
\cite{DMEref:B+E93,DMEref:ENS94,DMEref:BGE89}.
An alternative approach can be found in \cite{DMEref:BarSten00}, where the
case $\omega=0$, $\nu=0$ is treated, but it should be possible to use the
method for the general case as well.
\DMEendIND{Liouville operator!completeness of eigenvectors}
\DMEendIND{damping bases!completeness}

Note also that the time dependence in \DMEeq{29b} is solely carried by the
operator basis, not by the phase space function.
This is reminiscent of -- and in fact closely related to -- the
interaction-picture formalism of unitary quantum evolutions. 
Owing to the non-unitary terms in $\cL$, those proportional to the decay rate
$A$, the evolution of $\varrho_t$ is not unitary in \DMEeq{29b}.
Of course, one could also have a description in which the operator basis does
not change in time, but the phase space function does.
It then obeys a partial differential equation of the Fokker--Planck type.
Concerning these matters, the reader should consult the standard quantum
optics literature as well as special focus books such as \cite{DMEref:Risken}.

\subsection{Positivity Conservation}\label{DMEsec:3c}
\DMEbegIND{Liouville operator!conserves positivity}
Let us now return to the question that we left in limbo in
Sect.~\ref{DMEsec:1b}:
Does the master equation \DMEeq{10} preserve the positivity of $\varrho_t$?

Suppose that $\varrho_{t=0}$ is not a general operator but really a
statistical operator.
Then the coefficients $\beta_n^{(k)}$ in \DMEeq{29d} are such that the
right-hand side is non-negative for $t=0$ and has unit trace.
Since $\varrho_0^{(0)}$ is the steady state $\varrho^{(\mathrm{ss})}$ 
of \DMEeq{11}
and $\check{\varrho}_0^{(0)}=\DMEone$ is the identity, we have
\begin{equation}
 \DMEtr{\varrho_{n}^{(k)}}= \DMEtr{\check{\varrho}_0^{(0)}\varrho_{n}^{(k)}}
=\delta_{n,0}\delta_{k,0} 
\end{equation}
as a consequence of \DMEeq{22c}, and
\begin{equation}
\beta_n^{(k)} =\DMEtr{\check{\varrho}_n^{(k)}\varrho_{t=0}} 
\end{equation}
implies $\beta_0^{(0)}=\DMEtr{\varrho_{t=0}}=1$.
So, all terms in \DMEeq{29d} are traceless with the sole exception of the
time independent $n=0,k=0$ term, which has unit trace.
This demonstrates once more that the trace of $\varrho_t$ is conserved, as we
noted in Sect.~\ref{DMEsec:1b} already.

The time independent $n=0,k=0$ term is also the only one in  \DMEeq{29d} that
is non-negative by itself.
By contrast, the expectation values of $\varrho_{n}^{(k)}$ 
are both positive and negative if
$n>0,k=0$ and even complex if $k\neq0$.
Now, since $\varrho_{t=0}\geq0$, the  $n=0,k=0$ term clearly dominates 
all others for $t=0$, and then it dominates them even more for $t>0$
because the weight of $\varrho_0^{(0)}=\varrho^{(\mathrm{ss})}$ 
is constant in time while the other $\varrho_{n}^{(k)}$'s
have weight factors that decrease exponentially with $t$.
We can therefore safely infer that the master equation \DMEeq{10} conserves
the positivity of $\varrho_t$.
\DMEendIND{Liouville operator!conserves positivity}

\subsection{Lindblad Form of Liouville Operators}
The Liouville operator of (\ref{DMEeq:10}) can be rewritten as
\begin{eqnarray}
\label{DMEeq:31a}
\cL\varrho
&=&\I \omega[\varrho,a^{\dagger}a]
+\frac{1}{2}A(\nu+1)\left([a,\varrho a^{\dagger}]
+[a\varrho,a^{\dagger}]\right)\nonumber\\
&&\hphantom{\I \omega[\varrho,a^{\dagger}a]}
+\frac{1}{2}A\nu\left([a^{\dagger},\varrho a]
+[a^{\dagger}\varrho,a]\right)
\end{eqnarray}
and then it is an example of the so-called \DMEind{Lindblad form} of Liouville 
operators.\index{Liouville operator!Lindblad form} 
The general Lindblad form is
\begin{equation}
\label{DMEeq:31b}
\cL\varrho=\frac\I {\hbar}[\varrho,H]
+\sum_j\left([V_j^{\dagger},\varrho V_j]+[V_j^{\dagger}\varrho,V_j]\right)
\end{equation}
where $H=H^{\dagger}$ is the Hermitian Hamilton operator that generates the
unitary part of the evolution.
Clearly, any $\cL$ of this form will conserve the trace, but the requirement
of trace conservation would also be met if the sum were subtracted in
\DMEeq{31b} rather than added.
As Lindblad showed \cite{DMEref:Lindblad}, however, this option is only
apparent: all terms of the form 
$[V_j^{\dagger},\varrho V_j]+[V_j^{\dagger}\varrho,V_j]$ 
in a Liouville operator must come with a positive weight.
He further demonstrated that all $\cL$'s of the form \DMEeq{31b} surely
conserve the positivity of $\varrho$ provided that all the $V_j$'s are bounded.
This fact has become known as the 
\emph{Lindblad theorem}\index{Lindblad theorem}.
In the case that some $V_j$ is not bounded, positivity may be conserved or
not.
A proof of the Lindblad theorem is far beyond the scope of these lectures.
Reference \cite{DMEref:AliLen} is perhaps a good starting point for the reader
who wishes to learn more about these matters. 

We must in fact recognize that \DMEeq{31a} obtains for
\begin{equation}
 V_1=\sqrt{\thalf A(\nu+1)}\,a^{\dagger}\;,\quad V_2=\sqrt{\thalf A\nu}\,a
\end{equation}
in \DMEeq{31b} and these are actually \emph{not} bounded so that the Lindblad
theorem does not apply.
Fortunately, we have other arguments at hand, namely the ones of
Sect.~\ref{DMEsec:3c}, to show that the master equation \DMEeq{10} conserves
positivity. 

But while we are at it, let's just give a little demonstration of what goes
wrong when a master equation is not of the Lindblad form.
Take 
\begin{equation}
\label{DMEeq:31c}
\frac{\partial}{\partial t}\varrho_t
=VV^{\dagger}\varrho_t-2V^{\dagger}\varrho_t V+\varrho_t VV^{\dagger}\;,
\end{equation}
for example,
and assume that the initial state is pure, 
$\varrho_{t=0}=|\psi_0\rangle\langle \psi_0|$. 
Then we have at $t=\D t>0$
\begin{equation}
\varrho_{t=\D t}=\varrho_{t=0}
+\D t\frac{\partial\varrho_t}{\partial t}\bigg|_{t=0}
=|\psi_0\rangle\langle \psi_0|+\D t
\bigl(|\psi_2\rangle\langle \psi_0|-2|\psi_1\rangle\langle \psi_1|
+|\psi_0\rangle\langle \psi_2|\bigr)\;,
\end{equation}
where $|\psi_1\rangle=V^{\dagger}|\psi_0\rangle$
and $|\psi_2\rangle=VV^{\dagger}|\psi_0\rangle$.
Choose $|\psi_0\rangle$ such that $\langle \psi_1|\psi_0\rangle=0$
and  $\langle \psi_1|\psi_2\rangle=0$, and calculate the probability 
for $|\psi_1\rangle$ at time $t=\D t$,
\begin{equation}\label{DMEeq:31e}
\frac{\langle \psi_1|\varrho_{t=\D t}|\psi_1\rangle}
{\langle \psi_1|\psi_1\rangle}
=-2\D t<0\;!
\end{equation}
Positivity is violated already after an infinitesimal time step 
and, therefore, \DMEeq{31c} is not really a master equation.

\subsection*{Homework Assignments}
\addcontentsline{toc}{subsection}{\protect\numberline{}Homework Assignments}
\begin{enumerate}
\renewcommand{\labelenumi}{\textbf{\theenumi}}
\addtocounter{enumi}{\value{DMEhw}}
\item \label{DMEhw:3a}
Show the equivalence of the $\lambda$ and $\bar{\lambda}$ 
versions of \DMEeq{27a}.
\item 
If $F=f_1(Q;P)=f_2(P;Q)$, what is the relation 
between $f_1(Q',P')$ and $f_2(P',Q')$?
\item 
Consider $\Gamma=\thalf(QP+PQ)$, the generator of 
\DMEind{scaling transformations}.
For real numbers $\mu$,
write $\Exp{\I\mu\Gamma}$ in $Q,P$-ordered and $P,Q$-ordered form.
\item 
Find the Wigner function of the number operator $a\adj a$.
\item 
Construct an explicit (and simple!) example for \DMEeq{31c}--\DMEeq{31e},
that is: specify $V$ and $|\psi_0\rangle$.
\setcounter{DMEhw}{\value{enumi}}
\end{enumerate}


\section{Fourth Lecture: Quantum-Optical Applications}
Single atoms are routinely passed 
through high-quality resonators in experiments performed in Garching and Paris,
much like it is depicted in Fig.~\ref{DMEfig:1};
see the review articles 
\cite{DMEref:OAM/MPQa,DMEref:OAM/ENSa,DMEref:OAM/MPQb,DMEref:OAM/ENSb,%
DMEref:WVE02} 
and the references cited in them.
In a set-up typical of the Garching experiments, the atoms deposit energy
into the resonator and so compensate for the losses that result from
dissipation.
The intervals between the atoms are usually so large that an atom is long gone
before the next one comes, so that at any time at most one atom is inside the
cavity, all other cases being extremely rare.
And, therefore, the fitting name ``one-atom maser'' or ``micromaser'' 
\index{one-atom maser|DMEseealso{micromaser}}%
\index{micromaser|DMEseealso{one-atom maser}}%
has been coined for this system.

The properties of the steady state of the radiation field that is
established in the one-atom maser are determined by the values of 
several parameters of which the photon decay rate $A$, the thermal 
photon number $\nu$, and the atomic arrival rate $r$ are the most 
important ones.
Rare exceptions aside, an atom is entangled with the photon mode after
emerging from the cavity and it becomes entangled with the next atom 
after that has traversed the resonator.
As a consequence, measurements on the exiting atoms reveal intriguing
correlations which are the primary source of information of the photon 
state inside the cavity. 
A wealth of phenomena has been studied in these experiments over the last
10--15 years, and we look forward to seeing many more exciting results 
in the future.

\subsection{Periodically Driven Damped Oscillator}
To get an idea of the theoretical description of such experiments and to show
a simple, yet typical, application of the damping bases of 
Sect.~\ref{DMEsec:2} without, however, getting into too much realistic detail,
we'll consider the following somewhat idealized scenario.
The atoms come at the regularly spaced instants 
$t=\dots,-2T,-T,0,T,2T,\dots$ and each
atom effects a quasi-instantaneous change of the field of the cavity mode
($\equiv$\,the oscillator) that is specified by 
a ``\DMEind{kick operator}'' $\cK$ which is such that
\begin{equation}
  \label{DMEeq:32a}
  \varrho_{jT-0}\to\varrho_{jT+0}=(1+\cK)\varrho_{jT-0}
\end{equation}
states how $\varrho_t$ changes abruptly at $t=jT$ 
($j=0,\pm1,\pm2,\dots$).
Physically speaking, this abruptness just means that the time spent by an atom
inside the resonator is very short on the scale set by the decay constant $A$.

The evolution between  the kicks follows the master equation \DMEeq{10}.
We have 
\begin{equation}
\label{DMEeq:32b}
\frac{\partial}{\partial t}\varrho_t
=\cL\varrho_t+\cK\sum_j\delta(t-jT)\varrho_{t-0}
\end{equation}
as a formal master equation that incorporates the kicks \DMEeq{32a}.
The $\cL\varrho_t$ term accounts for the decay of the photon field in the
resonator.
In Sect.~\ref{DMEsec:1a} we derived the form of $\cL$ by pretending that the
dissipation is the result of a very weak interaction with very many atoms.
But, of course, this model must not be regarded as the true physical
mechanism.
Actually, the loss of electromagnetic energy is partly caused by leakage
through the cavity openings (through which the atoms enter and leave) and
partly by the ohmic resistance from which the currents suffer that are induced
on the surface of the conducting walls.
The ohmic losses are kept very small by fabricating the cavity from
superconducting metal (niobium below its critical temperature).

\begin{figure}[!t]
\centering
\includegraphics[bb=160 425 455 630]{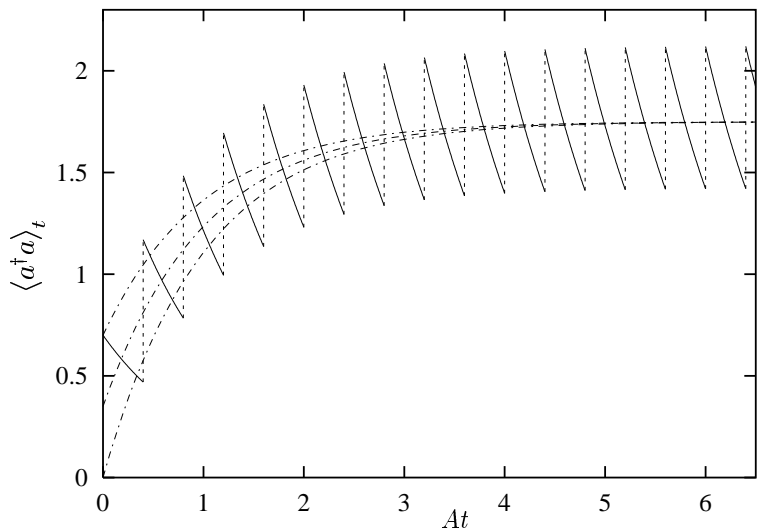}  
\caption[Mean number of excitations of a periodically kicked oscillator.]
{\label{DMEfig:4}%
Mean number of excitations of a periodically kicked oscillator; see text} 
\end{figure}

Thus, the term  $\cL\varrho_t$ in \DMEeq{32b} has nothing to do with the atoms
that the experimenter passes through the resonator in a micromaser experiment.
These atoms interact strongly with the photons and give rise to the kick
operator $\cK$.
Figure~\ref{DMEfig:4} shows what to expect under the circumstances to which
\DMEeq{32b} refers.
The mean number of excitations, $\DMEexpect{a\adj a}_t$, decays between the
kicks (solid $\frac{\qquad}{\qquad}$ line)
in accordance with \DMEeq{C2} and changes abruptly when a kick happens
(vertical dashed $---$ lines at $At=0.4,0.8,1.2,\dots$).
After an initial period, which lasts about a dozen kicks in
Fig.~\ref{DMEfig:4}, the 
\DMEind{cyclically steady state} $\varrho^{\mathrm{(css)}}_t$
is reached whose defining property is that it is the periodic solution of
\DMEeq{32b}, 
\begin{equation}
  \label{DMEeq:32c}
  \varrho_{t+T}^{\mathrm{(css)}}=\varrho_t^{\mathrm{(css)}}\;.
\end{equation}
Its value just before a kick is determined by
\begin{equation}
\varrho_{t=-0}^{\mathrm{(css)}}
=\Exp{\cL T}\varrho_{t=+0}^{\mathrm{(css)}}
=\Exp{\cL T}(1+\cK)\varrho_{t=-0}^{\mathrm{(css)}}\;,
\end{equation} 
and we have 
\begin{equation}
  \label{DMEeq:32e}
\varrho_{t}^{\mathrm{(css)}}
=\Exp{\cL t}\varrho_{t=+0}^{\mathrm{(css)}}  
=\Exp{-\cL(T-t)}\varrho_{t=-0}^{\mathrm{(css)}}  
\end{equation}
for $0<t<T$, that is: between two successive kicks.

In passing, it is worth noting that a recent experiment \cite{DMEref:VBWW00}, 
in which photon states of a definite photon number (Fock states) 
were prepared in a \DMEind{micromaser}, 
used a periodic scheme for pumping and probing.
The theoretical analysis \cite{DMEref:BEVW00} benefitted from damping-bases
techniques. 

The fine detail that we see in Fig.~\ref{DMEfig:4} is usually not of primary
interest, partly because experiments tend to not resolve it.
For example, if one asks how long it takes to reach the cyclically steady
state, all one needs to know is the time-averaged behavior of the smooth
dash-dotted $-\cdot-\cdot-$ lines in Fig.~\ref{DMEfig:4}.
In the cyclically steady state, the meaning of ``time-averaged'' is
hardly ambiguous, we simply have
\begin{equation}
  \label{DMEeq:33a}
\bar{\varrho}^{\mathrm{(ss)}}
=\frac{1}{T}\int_0^T\D t\,\varrho_{t}^{\mathrm{(css)}}  \;,
\end{equation}
where it does not matter over which time interval we average as long as it
covers one or more periods of $\varrho_{t}^{\mathrm{(css)}}$.
The periodicity \DMEeq{32c} of $\varrho_{t}^{\mathrm{(css)}}$ implies that
$\partial\varrho_{t}^{\mathrm{(css)}}/\partial t$ is zero on average and,
therefore, we obtain
\begin{equation}
\cL\bar{\varrho}^{\mathrm{(ss)}}
+\frac{1}{T}\cK \varrho_{t=-0}^{\mathrm{(css)}}=0 
\end{equation}
when time averaging \DMEeq{32b}.
When combined with what we get upon using \DMEeq{32e} in \DMEeq{33a},
\begin{equation}
\bar{\varrho}^{\mathrm{(ss)}}
=\frac{1-\Exp{-\cL T}}{\cL T}\varrho_{t=-0}^{\mathrm{(css)}}  \;,
\end{equation}
it yields the equation that determines $\bar{\varrho}^{\mathrm{(ss)}}$,
\begin{equation}
  \label{DMEeq:33d}
 \cL\bar{\varrho}^{\mathrm{(ss)}}
+\cK\frac{\cL}{1-\Exp{-\cL T}}\bar{\varrho}^{\mathrm{(ss)}}=0 \;.
\end{equation}
For $\cK=0$, it is of course solved by $\rho^{\mathrm{(ss)}}=\rho_0^{(0)}$
of \DMEeq{11}.

A master equation for the time-averaged evolution, that is: an equation obeyed
by the time-averaged statistical operator $\bar{\varrho}_t$, cannot be
\emph{derived} from \DMEeq{32b} for the same reasons for which one cannot
derive the macroscopic Maxwell equations from the microscopic ones.
But they can be \emph{inferred} with physical arguments that are more than
just reasonably convincing.
The task is actually easier here because we have to deal with temporal
averages only whereas one also needs spatial averages in the case of
electromagnetism. 

Imagine, then, that a linear time average is taken of \DMEeq{32b},
\begin{equation}
  \label{DMEeq:33e}
  \frac{\partial}{\partial t}\bar{\varrho}_t=\cL\bar{\varrho}_t
+\cK\,(\textbf{?})\,\bar{\varrho}_t\;,
\end{equation}
where $(\textbf{?})\,\bar{\varrho}_t$ is the ill-determined average of the
summation in \DMEeq{32b} that accounts for the periodic kicks \DMEeq{32a}.
We require, of course, that $\bar{\varrho}^{\mathrm{(ss)}}$ is the steady
state of \DMEeq{33e}.
In view of \DMEeq{33d}, this requirement settles the issue \cite{DMEref:B+E95},
\begin{equation}
  \label{DMEeq:33f}
  \frac{\partial}{\partial t}\bar{\varrho}_t=\cL\bar{\varrho}_t
+\cK\frac{\cL}{1-\Exp{-\cL T}}\bar{\varrho}_t\;,
\end{equation}
which we now accept as the master equation that describes the time-averaged
evolution.\index{master equation!for time-averaged evolution}
This is another case where an equation is ultimately justified by its
consequences.

We run a simple, but important consistency check on \DMEeq{33f}.
If the spacing $T$ between the atoms decreases, $T\to0$, and also the effect
of a single atom, $\cK=p\cM$ with $p\to0$, such that their ratio $r=p/T$ is
constant, then the situation should be equivalent to that of Poissonian
arrival statistics with rate $r$ and each atom effecting a kick $\cM$.
Indeed, $\DMEeq{33f}$ turns into
\begin{equation}
  \label{DMEeq:33g}
  \frac{\partial}{\partial t}\bar{\varrho}_t=\cL\bar{\varrho}_t
+r\cM\bar{\varrho}_t\;,  
\end{equation}
as it should, because this is the familiar \DMEind{Scully--Lamb equation} that
is known to apply in the case of \DMEind{Poissonian statistics}.
Thus, with $\cK=rT\cM$ in \DMEeq{33f} we obtain a master equation,
\begin{equation}
  \frac{\partial}{\partial t}\bar{\varrho}_t=\cL\bar{\varrho}_t
+r\cM\frac{\cL T}{1-\Exp{-\cL T}}\bar{\varrho}_t\;,
\end{equation}
that interpolates between the Poissonian Scully--Lamb limit of $T=0$ and that
of highly regular arrival times, $T=1/r$.

The damping bases associated with $\cL$ are the crucial tool for handling
\DMEeq{33f}.
We write
\begin{equation}
\bar{\varrho}_t=\sum_{n=0}^{\infty}\sum_{k=-\infty}^{\infty}
\bar{\alpha}_n^{(k)}(t)\varrho_n^{(k)}  
\end{equation}
and obtain differential equations for the numerical coefficients 
$\bar{\alpha}_n^{(k)}$,
\begin{equation}
  \bar{\alpha}_n^{(k)}(t)=\DMEtr{\check{\varrho}_n^{(k)}\bar{\varrho}_t}\;,
\end{equation}
by exploiting
\begin{equation}
  f(\cL)\varrho_n^{(k)}= f\bigl(\lambda_n^{(k)}\bigr)\varrho_n^{(k)}\;,
\qquad\lambda_n^{(k)}=-\I k\omega-(n+\thalf|k|)A\;,
\end{equation}
which holds for any function $f(\cL)$ simply because $\varrho_n^{(k)}$ is the
right eigenvector of $\cL$ to eigenvalue $\lambda_n^{(k)}$.
In this way, \DMEeq{33f} implies
\begin{equation}
  \label{DMEeq:34d}
\frac{\D}{\D t}\bar{\alpha}_n^{(k)}=\lambda_n^{(k)}\bar{\alpha}_n^{(k)}
+\sum_{n',k'}\cK_{n,n'}^{(k,k')}
\frac{\lambda_{n'}^{(k')}}
{1-\Exp{-\lambda_{n'}^{(k')}T}}\bar{\alpha}_{n'}^{(k')}
\end{equation}
where
\begin{equation}
\cK_{n,n'}^{(k,k')}=\DMEtr{\check{\varrho}_n^{(k)}\cK\varrho_{n'}^{(k')}}  
\end{equation}
is the matrix representation\index{kick operator!matrix representation} 
of the kick operator $\cK$ in the damping bases.
The seemingly troublesome ratio in \DMEeq{33f} is not a big deal anymore in
\DMEeq{34d}, where
\begin{equation}
  \left.\frac{\lambda}{1-\Exp{-\lambda T}}
\right|_{\lambda\to\lambda_0^{(0)}=0}
=\frac{1}{T}\;,
\end{equation}
of course.

Let us illustrate this for the particularly simple kick operator specified by
\begin{equation}
\cK\varrho=p\left(a^{\dagger}
\frac{1}{\sqrt{aa^{\dagger}}}
\varrho\frac{1}{\sqrt{aa^{\dagger}}}a-\varrho\right)  \qquad\mbox{with}\quad
0\leq p\leq1\;,
\end{equation}
which describes the over-idealized situation in which an atom adds one photon
with probability $p$ and does nothing with probability $1-p$.
Here,
\begin{equation}
  \cK_{n,n'}^{(k,k')}=\delta_{k,k'}\cK_{n,n'}^{(k,k)}
\end{equation}
so that the evolution does not mix $\bar{\alpha}_n^{(k)}$'s of different $k$
values.
As a further simplification it is therefore permissible to just consider the
$k=0$ terms.
For $\nu=0$ (homework assignment \ref{DMEhw:4a} deals with $\nu>0$), the
generating functions \DMEeq{17d} and \DMEeq{18a} give
\begin{eqnarray}
  \sum_{m,n=0}^{\infty}y^m\cK_{m,n}^{(0,0)}x^n&=&
\DMEtr{\left(1+y\right)\power{a^{\dagger}a}\cK\frac{1}{1+x}
\left(\frac{x}{1+x}\right)\power{a^{\dagger}a}}\nonumber\\
&=&\frac{py}{1-xy}=\sum_{n=0}^{\infty}py^{n+1}x^n\;,
\end{eqnarray}
where we let $\cK$ act to the left,
\begin{equation}
  (1+y)^{a^{\dagger}a}\cK=py(1+y)^{a^{\dagger}a}\;,
\end{equation}
and recall the trace evaluation of \DMEeq{19a}. 
We find
\begin{equation}
  \label{DMEeq:35e}
  \cK_{m,n}^{(0,0)}=p\delta_{m,n+1}\;,
\end{equation}
and the equation for $\bar{\alpha}_n^{(0)}(t)$ then has the explicit form
\begin{equation}
  \label{DMEeq:35f}
 \frac{\D}{\D t}\bar{\alpha}_n^{(0)}=-nA\bar{\alpha}_n^{(0)}
+p\frac{(n-1)A}{\Exp{(n-1)AT}-1}\bar{\alpha}_{n-1}^{(0)}\;. 
\end{equation}
This differential recurrence relation is solved successively by
\begin{eqnarray}
&&\bar{\alpha}_0^{(0)}(t)=\DMEtr{\bar{\varrho}_t}=1\;,\nonumber\\[1ex]
&&\bar{\alpha}_1^{(0)}(t)=\DMEexpect{a\adj a}_t=
\DMEexpect{a\adj a}_{\infty}
+\left(\DMEexpect{a\adj a}_0-\DMEexpect{a\adj a}_{\infty}\right)\Exp{-At}
\;,\nonumber\\[1ex]
&&\bar{\alpha}_2^{(0)}(t)=\frac{1}{2}\DMEexpect{{a\adj}^2 a^2}_t=
\frac{1}{2}\DMEexpect{{a\adj}^2 a^2}_{\infty}
+\frac{1}{2}\left(\DMEexpect{{a\adj}^2 a^2}_0
-\DMEexpect{{a\adj}^2 a^2}_{\infty}\right)\Exp{-2At}\nonumber\\
&&\hphantom{\bar{\alpha}_2^{(0)}(t)=\frac{1}{2}\DMEexpect{{a\adj}^2 a^2}_t=}
+\DMEexpect{{a\adj}^2 a^2}_{\infty}
\left(\!\frac{\DMEexpect{a\adj a}_0}{\DMEexpect{a\adj a}_{\infty}}-1\!\right)
\!\left(\!\Exp{-At}-\Exp{-2At}\right)
\end{eqnarray}
and so forth, where
\begin{equation}
 \DMEexpect{a\adj a}_{\infty}=\frac{p}{AT}\;,\quad
 \DMEexpect{{a\adj}^2 a^2}_{\infty}=\frac{p}{AT}\frac{p}{\Exp{AT}-1}\;,\quad
\ldots
\end{equation}
are the expectations values in the time-averaged steady state
$\bar{\varrho}^{\mathrm{(ss)}}$. 

Actually, Fig.~\ref{DMEfig:4} just shows this $\DMEexpect{a\adj a}_t$ 
for $p=0.7$ and $AT=0.4$,
so that $\DMEexpect{a\adj a}_{\infty}=1.75$ and the approach to this
asymptotic value is plotted for $\DMEexpect{a\adj a}_0=0,0.35,0.7$ by the
three dash-dotted $-\cdot-\cdot-$ curves.
The time-averaged value of $\DMEexpect{a\adj a}_t$ is not well defined at
$t=0$, the instant of the first kick, any value in the range $0\cdots0.7$ 
can be justified equally well.
The memory of this arbitrary initial value is always lost quickly.

\subsection{Conditional and Unconditional Evolution}\label{DMEsec:4b}
Let us now be more realistic about the effect an atom has on the photon state.
In fact, we have worked that out already in Sect.~\ref{DMEsec:1} for the case
of atoms incident in state $\UP$ or in state $\DN$ and a resonant
Jaynes--Cummings coupling between the photons and the atoms.
For the theoretical description of many one-atom maser experiments this is
actually quite accurate, and it will surely do for the purpose of these
lectures. 

\begin{figure}[!t]
\centering
\includegraphics{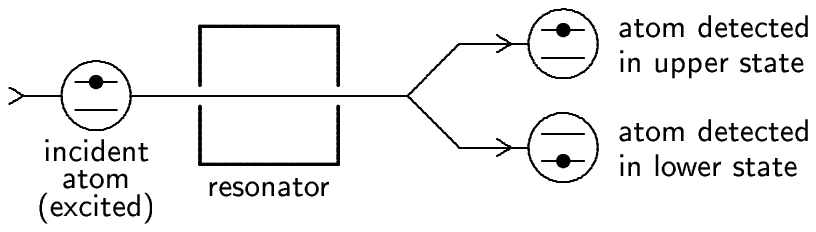}  
\caption[The final state of the exiting atom is detected.]
{\label{DMEfig:5}%
The final state of the exiting atom is detected: Is it $\UP$ or $\DN$\,?} 
\end{figure}

Since the atoms should deposit energy into the resonator as efficiently as
possible, they are prepared in the $\UP$ state.
The net effect of a single atom is then available in \DMEeq{B1},
which we now present conveniently as
\begin{equation}
  \label{DMEeq:36a}
  \cM\varrho_t=\cA\varrho_t+\cB\varrho_t-\varrho_t=(\cA+\cB-1)\varrho_t
\end{equation}
with
\begin{eqnarray}
  \label{DMEeq:36b}
\cA\varrho_t
&=&a\adj\frac{\sin\left(\phi\sqrt{aa^{\dagger}}\,\right)}{\sqrt{aa^{\dagger}}} 
\varrho_{t}
\frac{\sin\left(\phi\sqrt{aa^{\dagger}}\,\right)}{\sqrt{aa^{\dagger}}}a
\;,\nonumber\\[1ex]
\cB\varrho_t
&=&\cos\left(\phi\sqrt{aa^{\dagger}}\,\right)\varrho_t
\cos\left(\phi\sqrt{aa^{\dagger}}\,\right)\;.
\end{eqnarray}
As the derivation in Sect.~\ref{DMEsec:1} shows, the term $\cA\varrho_t$
corresponds to the atom emerging in state $\DN$, and likewise $\cB\varrho_t$
refers to $\UP$.
Accordingly, the respective probabilities 
for the final atom states $\DN$ and $\UP$ are
\begin{equation}
  \mathrm{prob}(\UP\to\DN)=\DMEtr{\cA\varrho_t}\;,\qquad
  \mathrm{prob}(\UP\to\UP)=\DMEtr{\cB\varrho_t}\;,
\end{equation}
and the probabilities $p_{\dn}$, $p_{\up}$ that the 
\DMEind{state-selective detection}
of Fig.~\ref{DMEfig:5} finds the atom in $\DN$ or $\UP$ are
\begin{equation}
 p_{\dn}=\eta_{\dn}\DMEtr{\cA\varrho_t}\;,\qquad
 p_{\up}=\eta_{\up}\DMEtr{\cB\varrho_t}\;,
\end{equation}
respectively, where $\eta_{\dn},\eta_{\up}$ are the detection efficiencies.

The effect on the photon state of an atom traversing the cavity at time $t$
can, therefore, be written as
\begin{equation}
  \label{DMEeq:37a}
  \varrho_t\to\eta_{\dn}\cA\varrho_t+\eta_{\up}\cB\varrho_t
    +\bigl[(1-\eta_{\dn})\cA\varrho_t+(1-\eta_{\up})\cB\varrho_t\bigr]
\end{equation}
where the three terms correspond to detecting the atom in state $\DN$,
detecting it in state $\UP$, and not detecting it at all.
The probability for the latter case is
\begin{equation}
  \label{DMEeq:37b}
  \mathrm{prob}(\mathrm{no\ click})
=\DMEtr{(1-\eta_{\dn})\cA\varrho_t+(1-\eta_{\up})\cB\varrho_t}
=1-\DMEtr{\cC\varrho_t}\;,
\end{equation}
where we recognize that
\begin{equation}
  \label{DMEeq:37c}
\mathrm{prob}(\UP\to\DN)+\mathrm{prob}(\UP\to\UP)
=\DMEtr{(\cA+\cB)\varrho_t}=1  
\end{equation}
and introduce the \index{click operator}\emph{click operator} $\cC$,
\begin{equation}
  \label{DMEeq:37d}
  \cC=\eta_{\dn}\cA+\eta_{\up}\cB\;.
\end{equation}

We take for granted that the atoms arrive with rate $r$ at statistically
independent instants (Poissonian arrival 
statistics\index{Poissonian statistics} once more).
The change of $\varrho_t$ brought about by a single \emph{undetected} atom
is
\begin{equation}
  \label{DMEeq:37e}
  \Delta\varrho_t\Big|_{\mathrm{undetected\ atom}}
=\frac{(\cA+\cB-\cC)\varrho_t}{1-\DMEtr{\cC\varrho_t}}-\varrho_t\;,
\end{equation}
where the numerator is just the third term of \DMEeq{37a} and the denominator
is its trace.
We multiply this with the probability that there is an atom between $t$ and
$t+\D t$, which is $r\D t$, and with the probability that the atom escapes
detection, which is given in \DMEeq{37b} and equal to the denominator in
\DMEeq{37e}, and so get
\begin{equation}
  \D t\frac{\partial\varrho_t}{\partial t}\Big|_{\mathrm{undetected\ atoms}}
=r\D t\bigl[(\cA+\cB-\cC)\varrho_t-\varrho_t
+\DMEtr{\cC\varrho_t}\varrho_t\bigr]\;.
\end{equation}
We combine it with \DMEeq{10},
\begin{equation}
  \label{DMEeq:38a}
  \frac{\partial\varrho_t}{\partial t}\Big|_{\mathrm{photon\ decay}}
=\cL\varrho_t\;,
\end{equation}
to arrive at
\begin{equation}
  \label{DMEeq:38b}
  \frac{\partial}{\partial t}\varrho_t=\bigl[\cL+r(\cA+\cB-1)\bigr]\varrho_t
-r\bigl[\cC-\DMEtr{\cC\varrho_t}\bigr]\varrho_t\;,
\end{equation}
the master equation that applies \emph{between detection events}.%
\index{master equation!nonlinear \DMEmain}%
\index{master equation!between detection events}
Owing to the term that involves the click probability $\DMEtr{\cC\varrho_t}$,
this is a \emph{nonlinear} master equation unless
$\eta_{\dn}=\eta_{\up}\equiv\eta$ when $\DMEtr{\cC\varrho_t}=\eta$ for all
$\varrho_t$. 
Fortunately, the nonlinearity is of a very mild form, since we can write
\DMEeq{38b} as
\begin{equation}
  \label{DMEeq:38c}
   \frac{\partial}{\partial t}\varrho_t
=\cL_{\eta}\varrho_t-\DMEtr{\cL_{\eta}\varrho_t}\varrho_t
\end{equation}
with the \emph{linear} operator $\cL_{\eta}$ given by
\begin{equation}
  \cL_{\eta}=\cL+r(\cA+\cB-1)-r\cC=\cL+r\cM-r\cC\;,
\end{equation}
and then solve it by
\begin{equation}
  \label{DMEeq:38e}
 \varrho_t=\frac{\Exp{\cL_{\eta}t}\varrho_{t=0}}
{\DMEtr{\Exp{\cL_{\eta}t}\varrho_{t=0}}}\;. 
\end{equation}
In effect, we can just ignore the second term in \DMEeq{38c}, evolve
$\varrho_t$ linearly from $\varrho_{t=0}$ to $\Exp{\cL_{\eta}}\varrho_{t=0}$,
and then normalize this to unit trace.
The normalization is necessary because $\cL_{\eta}$ by itself does not
conserve the trace, except for $\eta_{\dn}=\eta_{\up}=0$.
As we'll see in the fifth lecture, the normalizing denominator of \DMEeq{38e}
has a simple and important physical significance; see Sect.~\ref{DMEsec:5b}.

Note that there is a great difference between the situation in which the atoms
are not observed (you don't listen) and the situation in which they
are not detected (you listen but you don't hear anything).
The evolution of the photon field with unobserved atoms is the $\cC=0$ version
of \DMEeq{38b} that obtains for $\eta_{\dn}=\eta_{\up}=0$, which is just the
Scully--Lamb equation \DMEeq{33g} with $\cM$ of \DMEeq{36a}.
It describes the \emph{unconditional} evolution of the photon state.
By contrast, if the atoms are under observation but escape detection, the
nonlinear master equation \DMEeq{38b} or \DMEeq{38c} applies.
It describes the \emph{conditional} evolution of the photon state, conditioned
by the constraint that there are no detection events although detection is
attempted. 

The difference between conditional and unconditional evolution is perhaps best
illustrated in the extreme circumstance of perfectly efficient detectors,
$\eta_{\dn}=\eta_{\up}=1$, when no atom escapes detection.
Then \DMEeq{38b} turns into \DMEeq{38a} -- as it should because
``between detection events'' is tantamount to ``between atoms'' if every atom
is detected.
More generally, if the detection efficiency is the same for $\DN$ and $\UP$,
$\eta_{\dn}=\eta_{\up}=\eta$, so that each atom is detected with probability
$\eta$, we have
\begin{equation}
  \label{DMEeq:38f}
 \frac{\partial}{\partial t}\varrho_t=\bigl[\cL+(1-\eta)r\cM\bigr]\varrho_t  
\end{equation}
for the evolution between detection events.
This is the \DMEind{Scully--Lamb equation} 
with the actual rate $r$ replaced by the
effective rate $(1-\eta)r$, the rate of undetected atoms.

\subsection{Physical Significance of Statistical Operators}\label{DMEsec:4c}
\DMEbegIND{statistical operator!physical significance of \DMEmain{}s}
Master equation \DMEeq{38b} is the generalization of \DMEeq{38f} that takes
into account that $\DN$ atoms are not detected with the same efficiency as
$\UP$ atoms.
This asymmetry may originate in actually different detection devices or -- and
this is in fact the more important situation -- it is a consequence of the
question we are asking.

Assume, for example, that the $\DN$ detector clicked at $t=0$ and you want to
know how probable it is that the \emph{next} click of the $\DN$ detector
occurs between $t$ and $t+\D t$.
Clicks of the $\UP$ detector are of no interest to you whatsoever.
In an experiment you would just ignore them because they have no bearing on
your question.
The same deliberate ignorance\index{deliberate ignorance} enters the
theoretical treatment: you employ \DMEeq{38b} with $\eta_{\up}=0$ in the click
operator \DMEeq{37d} irrespective of the actual efficiency of the $\UP$
detector used in the experiment.
Likewise if your question were about the next click of the $\UP$ detector
you'd have to put $\eta_{\dn}=0$ in \DMEeq{37d}.

All of this is well in accord with the physical significance of the
statistical operator $\varrho_t$: it serves the sole purpose of enabling us to
make correct predictions about measurement at time $t$, in particular about
the probability that a certain outcome is obtained if a measurement is
performed. 
Such probabilities are always conditioned, they naturally depend on the
constraints to which they are subjected.
Therefore, it is quite possible that two persons have different statistical
operators for the same physical object because they take different conditions
into account.

Let us illustrate this point by a detection scheme that is simpler, and more
immediately transparent, than the standard one-atom maser experiment specified
by $\cA$ and $\cB$ of \DMEeq{36b}.
Instead we take 
\begin{eqnarray}
  \label{DMEeq:39a}
\cA\varrho_t
&=&\frac{1+(-1)\power{a\adj a}}{2}
\varrho_t
\frac{1+(-1)\power{a\adj a}}{2}
\;,\nonumber\\
\cB\varrho_t&=&\frac{1-(-1)\power{a\adj a}}{2}
\varrho_t\frac{1-(-1)\power{a\adj a}}{2}
\;,  
\end{eqnarray}
which can be realized by suitably prepared two-level atoms that have a
non-resonant interaction with the photon field and a suitable manipulation
prior to detecting $\DN$ or $\UP$ \cite{DMEref:ESW93}.
What is measured in such an experiment is the value of $(-1)\power{a\adj a}$,
the parity of the photon state.\index{parity measurement}
Detecting the atom in state $\DN$ indicates even parity, 
$(-1)\power{a\adj a}=1$, and a $\UP$ click indicates odd parity,
$(-1)\power{a\adj a}=-1$.

\begin{sidewaysfigure}
\centering
\includegraphics[height=320pt,bb=135 320 580 615]{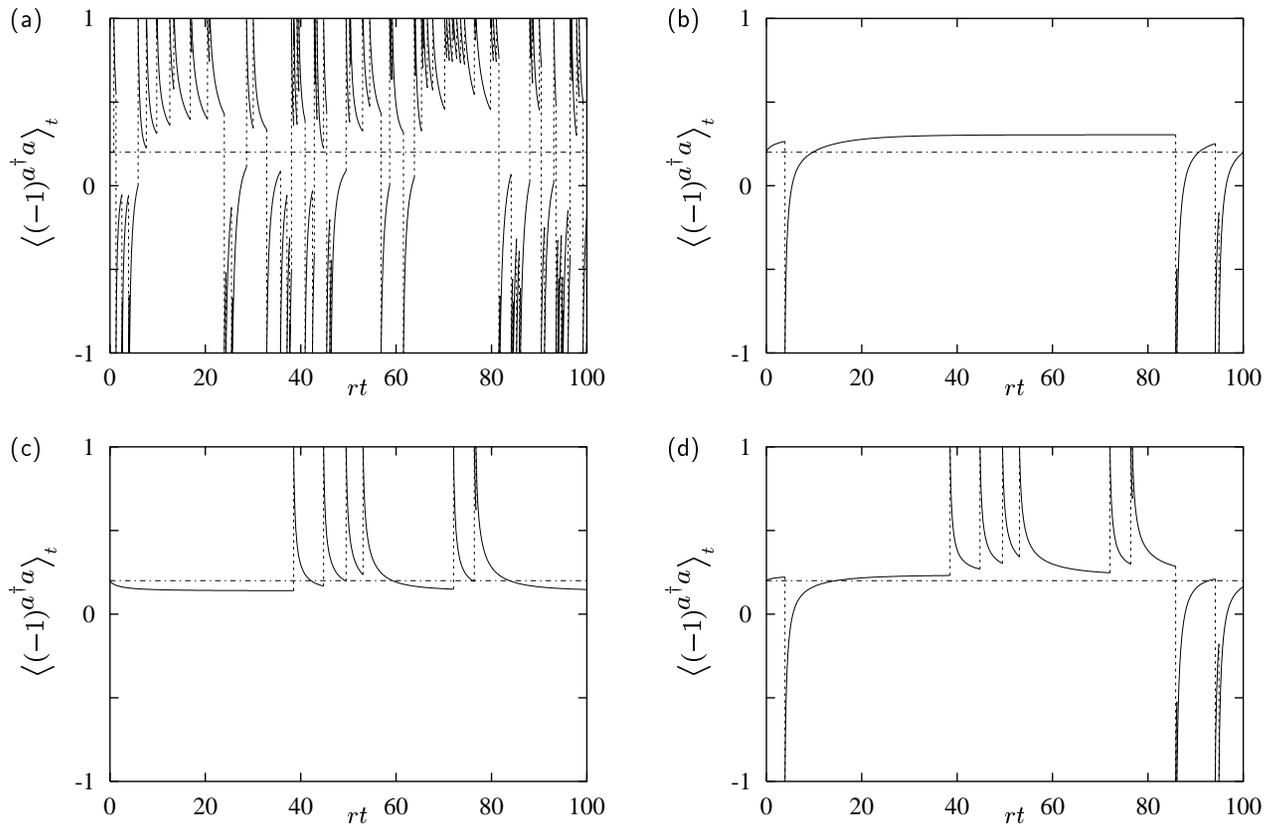}
\caption[Different, but equally consistent, statistical predictions (parity)]
{\label{DMEfig:6}%
Different, but equally consistent, statistical predictions about 
the same physical system: 
Parity expectation value when parity measurements are performed on an 
unpumped resonator; see text}
\end{sidewaysfigure}

\begin{sidewaysfigure}
\centering
\includegraphics[height=320pt,bb=135 320 575 615]{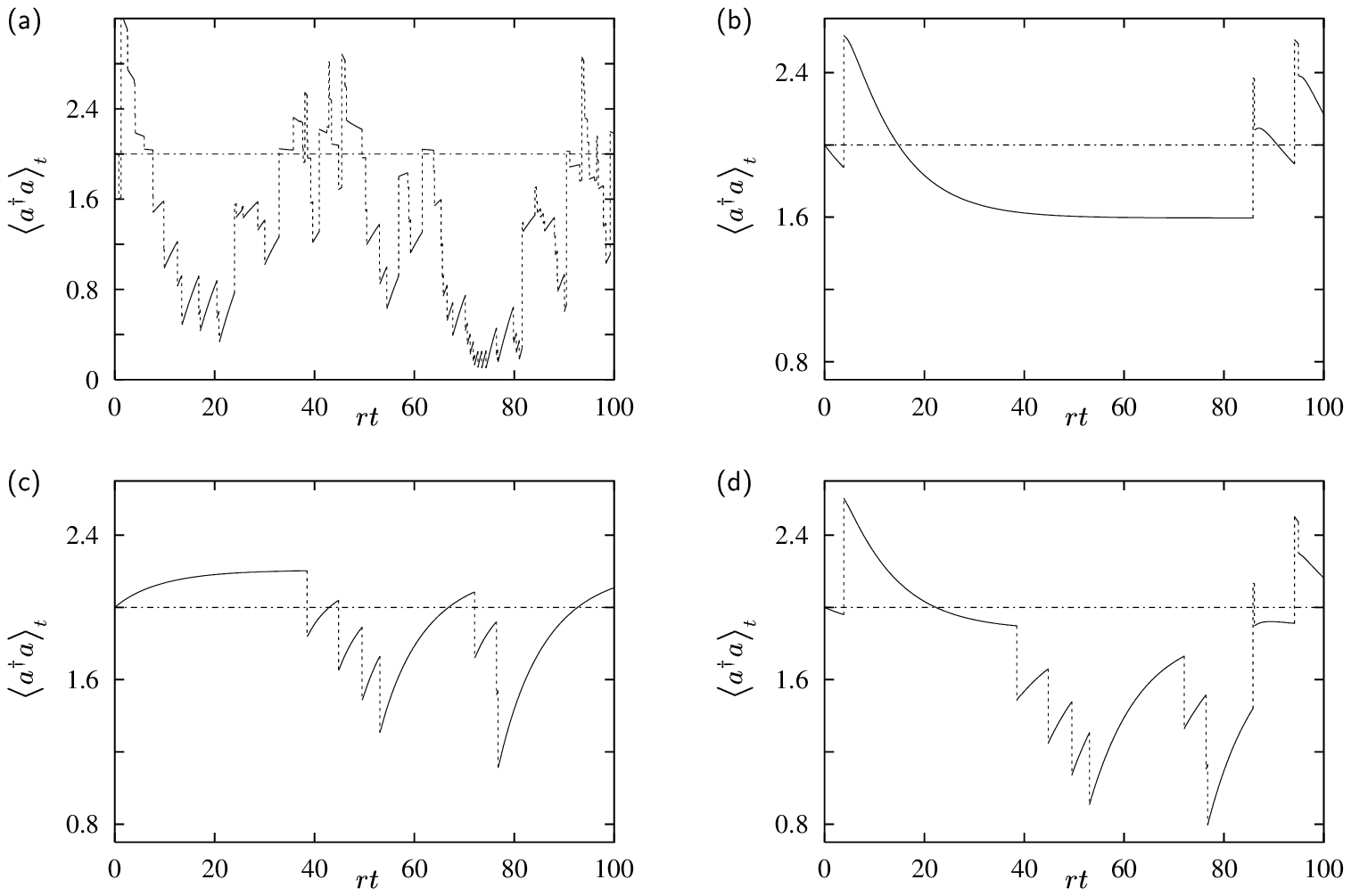}
\caption[Different, but equally consistent, statistical predictions (number)]
{\label{DMEfig:7}%
Different, but equally consistent, statistical predictions about 
the same physical system: 
Photon number expectation value when parity measurements are performed on an 
unpumped resonator; see text}
\end{sidewaysfigure}

Now consider the four cases of Figs.~\ref{DMEfig:6} and \ref{DMEfig:7}.
They refer to parity measurements on a \DMEind{one-atom maser} 
that is not pumped (no
resonant $\UP$ atoms are sent through) with $\nu=2$ (an atypically large
number of thermal photons for a micromaser experiment) and $r/A=10$.
The plots show the period $t=0\cdots100/r$ of the simulated experiment.
In Fig.~\ref{DMEfig:6} we see 
the parity expectation value as a function of $t$,
and in Fig.~\ref{DMEfig:7} we have the expectation value of the photon number.
The solid $\frac{\qquad}{}$ lines are the actual values, the vertical dashed
$---$ lines guide the eye through state-reduction jumps, and the horizontal
dash-dotted $-\cdot-\cdot-$ lines indicate the steady state values
\begin{equation}
  \DMEexpect{(-1)\power{a\adj a}}^{\mathrm{(ss)}}=\frac{1}{5}\;,\quad
  \DMEexpect{a\adj a}^{\mathrm{(ss)}}=2\;.
\end{equation}
On average, 100 atoms traverse the resonator in this time span, the actual
number is 108 here, of which 67 emerge in state $\DN$ (even parity) and 41 in
state $\UP$ (odd parity).
The final state is $\DN$ for 60\% of the atoms on average, and $\UP$ for 40\%.
The values chosen for the detection efficiencies are $\eta_{\dn}=10\%$ and
$\eta_{\up}=15\%$ so that each detector should register 6 atoms on average in
a period of this duration.
In fact, 7 $\DN$ clicks occurred 
(when $rt$=38.51, 44.80, 49.52, 53.07, 72.05, 76.41, and 76.75)
and 5 $\UP$ clicks
(when $rt$=3.88, 85.81, 86.09, 94.12, and 94.90).

Experimenter Bob pays no attention to the even-parity clicks of the $\DN$
detector, he is either not aware of them or has reasons to ignore them
deliberately.
Therefore he uses the nonlinear master equation \DMEeq{38c} 
with $\eta_{\dn}=0$ and $\eta_{\up}=0.15$ for the evolution between two 
successive clicks of the $\UP$ detector, and performs the 
\DMEind{state reduction}
\begin{equation}
  \label{DMEeq:39b}
  \varrho_t\to\frac{\cB\varrho_t}{\DMEtr{\cB\varrho_t}}
\end{equation}
whenever a $\UP$ click happens.
For example, to find the statistical operator $\varrho_{t=60/r}$ and then the
expectation values
\begin{equation}
  \label{DMEeq:39c}
  \textrm{Bob, } rt=60\,:\quad\DMEexpect{(-1)\power{a\adj a}}=0.1920\;,\quad
  \DMEexpect{a\adj a}=1.787\;,
\end{equation}
he applies \DMEeq{39b} to the $\UP$ click at $rt=3.88$, 
the only one between $rt=0$ and $rt=60$.
Bob's $\varrho_{t=0}$ is the steady state $\varrho^{\mathrm{(ss)}}$ of the
Scully--Lamb equation, consistent with his knowledge that the experiment has
been running long enough to have lost memory of its early history.
For \DMEeq{39a}, $\varrho^{\mathrm{(ss)}}$ is actually the thermal state
\DMEeq{11}, here with $\nu=2$.
Bob's detailed accounts are reported in Figs.~\ref{DMEfig:6}(b) and
\ref{DMEfig:7}(b).

Similarly, Figs.~\ref{DMEfig:6}(c) and \ref{DMEfig:7}(c) show what Chuck has
to say who pays no attention to $\UP$ clicks, but keeps a record of $\DN$
clicks. 
He uses \DMEeq{38c} with $\eta_{\dn}=0.1$ and $\eta_{\up}=0$ 
for the evolution between two successive $\DN$ clicks and performs the 
\DMEind{state reduction}
\begin{equation}
  \label{DMEeq:39d}
  \varrho_t\to\frac{\cA\varrho_t}{\DMEtr{\cA\varrho_t}}
\end{equation}
for each $\DN$ click.
To establish
\begin{equation}
  \label{DMEeq:39e}
  \textrm{Chuck, } rt=60\,:\quad\DMEexpect{(-1)\power{a\adj a}}=0.3042\;,\quad
  \DMEexpect{a\adj a}=1.599
\end{equation}
he has to do this for the four clicks prior to $t=60/r$.
Chuck uses the same $\varrho_{t=0}$ as Bob, of course, because both
have the same information about the preparation.

And then there is Doris who pays full attention to all detector clicks.
She uses \DMEeq{38c} with $\eta_{\dn}=0.1$ and $\eta_{\up}=0.15$ between
successive clicks, does \DMEeq{39d} for $\DN$ clicks and \DMEeq{39b} for $\UP$
clicks, and arrives at 
\begin{equation}
  \label{DMEeq:39f}
  \textrm{Doris, } rt=60\,:\quad\DMEexpect{(-1)\power{a\adj a}}=0.2995\;,\quad
  \DMEexpect{a\adj a}=1.390\;.
\end{equation}
Her account is shown in Figs.~\ref{DMEfig:6}(d) and \ref{DMEfig:7}(d).

For the sake of completeness, we also have  Figs.~\ref{DMEfig:6}(a) and
\ref{DMEfig:7}(a), where state reductions are performed for each atom, whether
detected or not, and the $\eta_{\dn}=\eta_{\up}=0$ version of \DMEeq{38c}
-- the ``between atoms'' equation \DMEeq{38a} --
applies between the reductions.  
What is obtained in this manner is of no consequence, however, because it
incorporates data that are never actually available.

Why, then, do we show Figs.~\ref{DMEfig:6}(a) and \ref{DMEfig:7}(a) at all?
Because one might think that they report the ``true state of affairs'' so that
\begin{equation}
  \textrm{all atoms, } rt=60\,:\quad
\DMEexpect{(-1)\power{a\adj a}}=0.4818\;,\quad  
\DMEexpect{a\adj a}=1.189
\end{equation}
would be the ``true expectation values'' of $(-1)\power{a\adj a}$ and 
$a\adj a$ at $t=60/r$.
And then one would conclude that 
the accounts given by Bob, Chuck, and Doris are wrong in some sense.
In fact, all three give \emph{correct}, though differing accounts, and the
various predictions for $t=60/r$ in \DMEeq{39c},  \DMEeq{39e}, and \DMEeq{39f}
are all statistically correct.
For, if you repeat the experiment very often you'll find that Bob's
expectation values are confirmed by the data, and so are Chuck's, and so are
Doris's.

But, of course, when extracting $\DMEexpect{a\adj a}_{t=60/r}$, say, 
from the data of the very many runs, you must take different subensembles for
checking Bob's predictions, or Chuck's, or Doris's. 
Bob's prediction \DMEeq{39c} refers to the subensemble characterized by a
single $\UP$ click at $rt=3.88$ and no other $\UP$ click between $rt=0$ and
$rt=60$, but any number of $\DN$ clicks.
Likewise, Chuck's prediction \DMEeq{39e} is about the subensemble that has
$\DN$ clicks at $rt$=38.51, 44.80, 49.52, and 53.07, no other $\DN$ clicks
and any number of $\UP$ clicks.
And Doris's subensemble is specified by having this one $\UP$ click, 
these four $\DN$ clicks, and no other clicks of either kind.
By contrast, no experimentally identifiable ensemble corresponds to
Figs.~\ref{DMEfig:6}(a) and \ref{DMEfig:7}(a); they represent sheer
imagination, not phenomenological reality.

In summary, although it is the same physical object (the privileged 
mode of the resonator) that Bob, Chuck, and Doris make predictions about, they
regard it as a representative of different ensembles, which are respectively
characterized by the information taken into account.
This illustrates the basic fact that a statistical operator is just an
encoding of what we know about the system.
Depending on the question we are asking, we may even have to ignore some of
the information deliberately.
The appropriate $\varrho_t$ is the one that pays due attention to the
pertinent conditions under which we wish to make statistical predictions.
Different conditions simply require different statistical operators.
That there are several $\varrho_t$'s for the same physical object is then not
bewildering, but rather expected. 

The ensembles that Bob, Chuck, and Doris make statements about 
need not be, and usually are not, real ensembles created by many 
repeated runs of the
experiment (or, since the system is ergodic 
\cite{DMEref:BESW94,DMEref:KueMaa01}, perhaps a single run of 
very long duration).
Rather, they are Gibbs ensembles in the standard meaning of statistical
mechanics: imagined ensembles characterized by the respective constraints and
consistent weights. 
The constraints are duly taken into account 
by the nonlinear master equation \DMEeq{38c}.

We should also mention that Bob's $\varrho_t$ is consistent with Chuck's and
Doris's by construction.
If we didn't know how they arrive at their respective statistical operators
we might wonder how we could verify that the three $\varrho_t$'s do not
contradict each other.
Consult \cite{DMEref:Mermin01a,DMEref:Mermin01b} 
if you find the question interesting. 
\DMEendIND{statistical operator!physical significance of \DMEmain{}s}

\subsection*{Homework Assignments}
\addcontentsline{toc}{subsection}{\protect\numberline{}Homework Assignments}
\begin{enumerate}
\renewcommand{\labelenumi}{\textbf{\theenumi}}
\addtocounter{enumi}{\value{DMEhw}}
\item \label{DMEhw:4a} 
Show that
\begin{equation}
  \sum_{m=0}^{\infty}y^m\check{\varrho}_m^{(0)}
=\frac{1+\nu}{1+\nu+\nu y}\left(1+\frac{y}{1+\nu+\nu y}\right)\power{a\adj a}
\end{equation}
is the $\nu>0$ generalization of \DMEeq{18a}, then use this and \DMEeq{17c} to
find the $\nu>0$ version of \DMEeq{35e}.
\item 
State the $\nu>0$ version of \DMEeq{35f} and solve the equations for
$n=0,1,2$.
\item 
Take the \DMEind{Scully--Lamb limit} 
of \DMEeq{35f}, that is: $T\to0$ after putting
$p=rT$, and find the steady state values of all $\bar{\alpha}_n^{(0)}$'s.
\item 
Evaluate $\displaystyle\bar{\varrho}^{\mathrm{(ss)}}%
=\sum_{n=0}^{\infty}\bar{\alpha}_n^{(0)}\varrho_n^{(0)}$
for these $\bar{\alpha}_n^{(0)}$'s. 
You should obtain
\index{Liouville operator!right eigenvectors!generating function}
\index{damping bases!generating function}%
\index{generating function!for the damping bases}%
\begin{equation}
 \bar{\varrho}^{\mathrm{(ss)}}
=\frac{1}{\nu+1}\left(\frac{\nu}{\nu+1}\right)\power{a\adj a}
\mathrm{L}_{\mbox{\footnotesize$a\adj a$}}^{(0)}\left(-\frac{r}{\nu A}\right)
\Exp{-r/A}\;. 
\end{equation}
In which sense is this yet another generating function for the
$\varrho_n^{(0)}$'s?
What do you get for $\nu\to0$?
\item 
Verify \DMEeq{37c} for $\cB$ of \DMEeq{36b}.
\item 
Show that
\begin{equation}
  \cM f(a\adj a)
=\left[\sin\left(\phi\sqrt{a\adj a}\,\right)\right]^2f(a\adj a-1)
-\left[\sin\left(\phi\sqrt{a\adj a+1}\,\right)\right]^2f(a\adj a)
\end{equation}
for $\cM$ of \DMEeq{36a}, then combine this with \DMEeq{C5} to find the steady
state of the \DMEind{Scully--Lamb equation} \DMEeq{33g}.
You should get
\begin{equation}
  \varrho^{\mathrm{(ss)}}=N\prod_{n=1}\power{a\adj a}\left[\frac{\nu}{\nu+1}
+\frac{r/A}{\nu+1}\left(\frac{\sin\left(\phi\sqrt{n}\,\right)}{\sqrt{n}}
\right)^2\right]
\end{equation}
as the steady state of the one-atom maser.%
\index{one-atom maser!steady state of the \DMEmain}
What is the physical significance of the normalization factor $N$?
How do you determine it?
\item 
Differentiate $\varrho_t$ of \DMEeq{38e} to verify that it solves \DMEeq{38c}.
\item 
In addition to Bob, Chuck, and Doris, there is also Alice who pays attention
to all detector clicks but doesn't care which detector fires.
Which version of \DMEeq{38c} does she employ, and how does she go about state
reduction?  
\item 
(a) Since you found Sect.~\ref{DMEsec:4c} very puzzling, read it again.\\
(b) If you are still puzzled, repeat (a), else proceed with (c).\\
(c) Convince your favorite skeptical colleague that there can be different,
but equally consistent, statistical operators for the same object.
\setcounter{DMEhw}{\value{enumi}}
\end{enumerate}


\section{Fifth Lecture: Statistics of Detected Atoms}
A one-atom maser is operated under steady-state conditions.
What is the probability to detect a $\DN$ atom between $t$ and $t+\D t$?
It is 
\begin{equation}
  \label{DMEeq:41a}
  r\D t\, \eta_{\dn}\DMEtr{\cA\varrho^{\mathrm{(ss)}}}\;,
\end{equation}
the product of
the probability $r\D t$ of having an atom in this time interval
($r>0$ is taken for granted from here on) 
and the probability $\eta_{\dn}\DMEtr{\cA\varrho^{\mathrm{(ss)}}}$
that the $\DN$ detector clicks if there is an atom.
Similarly,
\begin{equation}
  \label{DMEeq:41b}
  r\D t\, \eta_{\up}\DMEtr{\cB\varrho^{\mathrm{(ss)}}}  
\end{equation}
is the probability for a $\UP$ click between $t$ and $t+\D t$.
What multiplies $r\D t$ here are the traces of the first and the second term
of \DMEeq{37a}, respectively.
The third term gave us the no-click probability \DMEeq{37b}.  

The a~priori probabilities \DMEeq{41a} and \DMEeq{41b} make no reference to
other detection events.
But the detector clicks are not statistically independent.
For the example of the parity measurements of Sect.~\ref{DMEsec:4c},
you can see this very clearly in Fig.~\ref{DMEfig:6}(d), where the even-parity
clicks and the odd-parity clicks come in bunches.
This bunching is easily understood:
a $\DN$ click is accompanied by the state reduction \DMEeq{39d} so that
$\varrho_t$ is an even-parity state immediately after a $\DN$ click and,
therefore, the next atom is much more likely to encounter even parity than odd
parity.
The detection of the first atom conditions the probabilities for the second
atom, and for all subsequent ones as well.

\subsection{Correlation Functions}
\DMEbegIND{correlation functions}
Perhaps the simplest question one can ask in this context is: Given that a
$\DN$ atom was detected at $t=0$, what is now the probability to detect a
$\UP$ atom between $t$ and $t+\D t$?
Since detection events that occur earlier than $t$ are not relevant here, we
must use \DMEeq{38c} with $\eta_{\dn}=\eta_{\up}=0$ to propagate
$\varrho_{t=0}$ to $\varrho_t$,
\begin{equation}
  \label{DMEeq:41c}
  \varrho_t=\Exp{\cL_0 t}\varrho_{t=0}
\end{equation}
where
\begin{equation}
  \cL_0=\cL_{\eta}\Big|_{\eta_{\srdn}=\eta_{\srup}=0}=\cL+r(\cA+\cB-1)
\end{equation}
is the Liouville operator of the Scully--Lamb equation%
\index{Liouville operator!of the Scully--Lamb equation}%
\index{Scully--Lamb equation!Liouville operator of \DMEmain} 
that determines the steady state,
\begin{equation}
  \cL_0\varrho^{\mathrm{(ss)}}=0\;,\qquad
\lim_{t\to\infty}\Exp{\cL_0 t}\varrho_{t=0}=\varrho^{\mathrm{(ss)}}\;.
\end{equation}
The denominator of \DMEeq{38e} equals unity here because $\cL_0$ conserves the
trace,
\begin{equation}
  \DMEtr{\Exp{\cL_0 t}\varrho_{t=0}}=\DMEtr{\varrho_{t=0}}=1\;.
\end{equation}
With
\begin{equation}
  \varrho_{t=0}=\frac{\cA\varrho^{\mathrm{(ss)}}}
                     {\DMEtr{\cA\varrho^{\mathrm{(ss)}}}}
\end{equation}
accounting for the initial $\DN$ click at $t=0$ (which, under steady-state
conditions, is really any instant), the probability for a $\UP$ click at
$t\cdots t+\D t$ is then
\begin{equation}
  \label{DMEeq:41f}
  r\D t\,\eta_{\up}\DMEtr{\cB\Exp{\cL_0 t}\varrho_{t=0}}
= r\D t\,\frac{\DMEtr{\cB\Exp{\cL_0 t}\cA\varrho^{\mathrm{(ss)}}}}
                     {\DMEtr{\cA\varrho^{\mathrm{(ss)}}}}\;.
\end{equation}
If $t$ is very late, the memory of the initial state gets lost and \DMEeq{41f}
becomes equal to the a~priori probability \DMEeq{41b}.
The ratio of the two probabilities,
\begin{equation}
  \label{DMEeq:41g}
  G_{\dn\up}(t)=\frac{\DMEtr{\cB\Exp{\cL_0 t}\cA\varrho^{\mathrm{(ss)}}}}
   {\DMEtr{\cB\varrho^{\mathrm{(ss)}}}\DMEtr{\cA\varrho^{\mathrm{(ss)}}}}\;,
\end{equation}
is the correlation function for $\UP$ clicks after $\DN$ clicks.
There are no correlations if ${G_{\dn\up}=1}$, positive correlations if
${G_{\dn\up}>1}$, negative correlations if ${G_{\dn\up}<1}$.
The bunching of Fig.~\ref{DMEfig:6}(d) shows strong negative $\DN\UP$
correlations (odd after even) at short times.

Likewise, if we ask about $\DN$ clicks after $\UP$ clicks, we get
\begin{equation}
  G_{\up\dn}(t)=\frac{\DMEtr{\cA\Exp{\cL_0 t}\cB\varrho^{\mathrm{(ss)}}}}
   {\DMEtr{\cA\varrho^{\mathrm{(ss)}}}\DMEtr{\cB\varrho^{\mathrm{(ss)}}}}\;,
\end{equation}
and we have
\begin{equation}
  \label{DMEeq:41i}
  G_{\dn\dn}(t)=\frac{\DMEtr{\cA\Exp{\cL_0 t}\cA\varrho^{\mathrm{(ss)}}}}
   {\left[\DMEtr{\cA\varrho^{\mathrm{(ss)}}}\right]^2}\;, 
\qquad
  G_{\up\up}(t)=\frac{\DMEtr{\cB\Exp{\cL_0 t}\cB\varrho^{\mathrm{(ss)}}}}
   {\left[\DMEtr{\cB\varrho^{\mathrm{(ss)}}}\right]^2}\;, 
\end{equation}
for the correlation functions of clicks of the same kind.
Note that the detection efficiencies $\eta_{\dn},\eta_{\up}$ do not enter the
correlation functions \DMEeq{41g}--\DMEeq{41i}.
This is a result of normalizing the conditional probabilities to the
unconditional ones.
But see homework assignment \ref{DMEhw:5a}.

\begin{figure}[!t]
\centering
\includegraphics[bb=160 420 455 630]{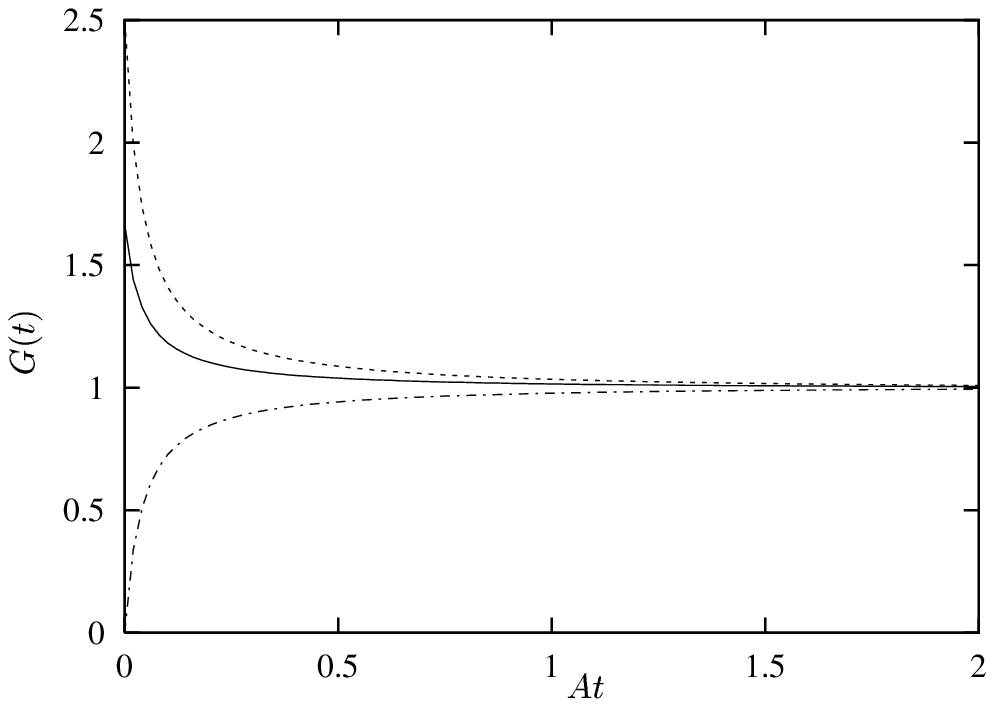}
\caption[Correlation functions for parity measurements]{\label{DMEfig:8}%
Correlation functions for parity measurements on an unpumped resonator.
The solid $\frac{\qquad}{}$ line shows the even-even correlation function 
$G_{\dn\dn}(t)$, the dashed $---$ line the odd-odd correlation function
$G_{\up\up}(t)$, and the dash-dotted $-\cdot-\cdot-$ line the cross
correlations $G_{\dn\up}(t)= G_{\up\dn}(t)$.
The plot is for $\nu=2$ in the expressions given 
in \DMEeq{42e} and \DMEeq{42f}}
\end{figure}

As an example we take once more the parity measurements of 
Figs.~\ref{DMEfig:6} and \ref{DMEfig:7}, for which $\cA,\cB$ are given in
\DMEeq{39a} so that $\cL_0=\cL$.
Then $\varrho^{\mathrm{(ss)}}$ is the thermal state of~\DMEeq{11} and the 
a~priori probabilities are    
\begin{eqnarray}
  \label{DMEeq:42a}
  \DMEtr{\cA\varrho^{\mathrm{(ss)}}}&=&
\dhalf+\dhalf\DMEexpect{(-1)\power{a\adj a}}^{\mathrm{(ss)}}
=\frac{\nu+1}{2\nu+1}\;,
\nonumber\\
  \DMEtr{\cB\varrho^{\mathrm{(ss)}}}&=&
\dhalf-\dhalf\DMEexpect{(-1)\power{a\adj a}}^{\mathrm{(ss)}}
=\frac{\nu}{2\nu+1}\;.
\end{eqnarray}
For $\nu=0$, there are no odd-parity clicks of the $\UP$ detector and,
therefore, we'll only consider the case of $\nu>0$.
Since
\begin{equation}
  \left.
    \begin{array}{l}
\cA\varrho^{\mathrm{(ss)}}\\[1ex] \cB\varrho^{\mathrm{(ss)}}
    \end{array}\right\}=\dhalf\varrho^{\mathrm{(ss)}}
\pm\dhalf\frac{1}{\nu+1}\left(-\frac{\nu}{\nu+1}\right)\power{a\adj a}\;,
\end{equation}
we have
\begin{equation}
  \Exp{\cL_0 t}(\cA+\cB)\varrho^{\mathrm{(ss)}}=\varrho^{\mathrm{(ss)}}
\end{equation}
and
\begin{equation}
  \Exp{\cL_0 t}(\cA-\cB)\varrho^{\mathrm{(ss)}}
=\frac{1}{2\nu+1}\lambda(t)\bigl[1-\lambda(t)\bigr]\power{a\adj a}\;, 
\end{equation}
where $\lambda(t)$ is given in \DMEeq{15} with $\lambda(0)=(2\nu+1)/(\nu+1)$.
The numerators of \DMEeq{41g}--\DMEeq{41i} are then easily evaluated and we
obtain
\begin{equation}
  \label{DMEeq:42e}
  G_{\dn\up}(t)= G_{\up\dn}(t)
=1-\left[1+(2\nu+1)^2(\Exp{At}-1)\right]^{-1}
\end{equation}
and
\begin{eqnarray}
  \label{DMEeq:42f}
 G_{\dn\dn}(t)&=&
1+\frac{\nu}{\nu+1}\left[1+(2\nu+1)^2(\Exp{At}-1)\right]^{-1}
\;,\nonumber\\[1ex]
G_{\up\up}(t)&=&
1+\frac{\nu+1}{\nu}\left[1+(2\nu+1)^2(\Exp{At}-1)\right]^{-1}
\;.
\end{eqnarray}
The cross correlations of \DMEeq{42e} vanish at $t=0$ (no even-parity click
immediately after an odd-parity click and vice versa) and increase
monotonically toward $1$.
The two same-click correlation functions of \DMEeq{42f} are always larger than
$1$ and decrease monotonically from their $t=0$ values
\begin{equation}
  \label{DMEeq:42g}
   G_{\dn\dn}(t=0)=\frac{2\nu+1}{\nu+1}\;,\qquad
   G_{\up\up}(t=0)=\frac{2\nu+1}{\nu}\;,
\end{equation}
which exceed unity and so confirm the bunching observed in
Fig.~\ref{DMEfig:6}(d).
For $\nu=2$, the parameter value of Figs.~\ref{DMEfig:6} and \ref{DMEfig:7},
the correlation functions are plotted in Fig.~\ref{DMEfig:8}.

Data of actual measurements of correlation functions for atoms emerging
from a real-life micromaser are reported in \cite{DMEref:ELBVWW98},
for example.\index{micromaser!measured correlation functions}
Theoretical values for some related quantities, such as the mean number of
successive detector clicks of the same kind, agree very well with the
experimental findings.
\DMEendIND{correlation functions}

\subsection{Waiting Time Statistics}\label{DMEsec:5b}
\DMEbegIND{waiting time statistics}
Here is a different question: 
A $\UP$ click happened at $t=0$, what is the probability that the next
$\DN$ clicks occurs between $t$ and $t+\D t$?
In marked contrast to the question asked before \DMEeq{41c}, we are now not
interested in any later click, but in the \emph{next} click, and this just
says that there are no other $\DN$ clicks before $t$.
Since we ignore deliberately all $\UP$ clicks at intermediate times,
we have to use \DMEeq{38c} with $\eta_{\up}=0$ in the click
operator $\cC$ of \DMEeq{37d}.

We introduce the following quantities:
\begin{equation}
  \begin{array}[b]{rcl}
r\D t\,p_{\dn}(t)
&=&\mbox{probability for a $\DN$ click at $t\cdots t+\D t$}\;,
\\[1ex]
p_{\mathrm{no}\dn}(t)&=&
\mbox{probability for no $\DN$ click before $t$}\;,
\\[1ex]
\D t\,P_{\mathrm{next}\dn}(t)&=&
\mbox{probability for the next $\DN$ click to happen at $t\cdots t+\D t$}\;. 
  \end{array}
\end{equation}
Since $1-r\D t\,p_{\dn}(t)$ is then the probability that there is no $\DN$
click between $t$ and $t+\D t$, we have
\begin{eqnarray}
  \label{DMEeq:43b}
&&p_{\mathrm{no}\dn}(t+\D t)
=p_{\mathrm{no}\dn}(t)\bigl[1-r\D t\,p_{\dn}(t)\bigr]
\nonumber\\[1ex]\mbox{or}&&
p_{\dn}(t)=-\frac{1}{r}\frac{\D}{\D t}\ln p_{\mathrm{no}\dn}(t)\;,  
\end{eqnarray}
and
\begin{eqnarray}
  \label{DMEeq:43c}
&& p_{\mathrm{no}\dn}(t+\D t)
= p_{\mathrm{no}\dn}(t)-\D t\,P_{\mathrm{next}\dn}(t)
\nonumber\\[1ex]\mbox{or}&&
P_{\mathrm{next}\dn}(t)=-\frac{\D}{\D t} p_{\mathrm{no}\dn}(t)
\end{eqnarray}
is another immediate consequence of the significance given to these quantities.

We know $p_{\dn}(t)$ from \DMEeq{37b} and \DMEeq{38e},
\begin{equation}
  \label{DMEeq:44a}
  p_{\dn}(t)=\frac{\DMEtr{\cC\Exp{\cL_{\eta}t}\varrho_{t=0}}}
                  {\DMEtr{\Exp{\cL_{\eta}t}\varrho_{t=0}}}
\end{equation}
where
\begin{equation}
  \cC=\eta_{\dn}\cA\qquad\mbox{and}\qquad
\varrho_{t=0}=\frac{\cB\varrho^{\mathrm{(ss)}}}
                   {\DMEtr{\cB\varrho^{\mathrm{(ss)}}}}
\end{equation}
in the present context.
As required by \DMEeq{43b}, the right-hand side of \DMEeq{44a} must be a
logarithmic derivative and, indeed, it is
because the identity
\begin{eqnarray}
  \label{DMEeq:44c}
  \frac{\D}{\D t}\DMEtr{\Exp{\cL_{\eta}t}\varrho_{t=0}}
&=&\DMEtr{\cL_{\eta}\Exp{\cL_{\eta}t}\varrho_{t=0}}
\nonumber\\[1ex]
&=&\DMEtr{(\cL_0-r\cC)\Exp{\cL_{\eta}t}\varrho_{t=0}}
\nonumber\\[1ex]
&=&-r\DMEtr{\cC\Exp{\cL_{\eta}t}\varrho_{t=0}}\;,
\end{eqnarray}
which uses the trace-conserving property of $\cL_0$, implies
\begin{equation}
  p_{\dn}(t)=-\frac{1}{r}\frac{\D}{\D t}
\ln\DMEtr{\Exp{\cL_{\eta}t}\varrho_{t=0}}\;.
\end{equation}
It follows that
\begin{equation}
  \label{DMEeq:44e}
 p_{\mathrm{no}\dn}(t)= \DMEtr{\Exp{\cL_{\eta}t}\varrho_{t=0}}\;,
\end{equation}
and then \DMEeq{43c}, \DMEeq{44c} give
\begin{equation}
  \label{DMEeq:44f}
  P_{\mathrm{next}\dn}(t)=r\DMEtr{\cC\Exp{\cL_{\eta}t}\varrho_{t=0}}\;.
\end{equation}
The ``important physical significance'' of the denominator in \DMEeq{38c} that
was left in limbo in Sect.~\ref{DMEsec:4b} is finally revealed in \DMEeq{44e}:
it is the probability that no atom is detected before~$t$.
Since an atom is surely detected if we just wait long enough, the limit
\begin{equation}
  \label{DMEeq:44g}
  p_{\mathrm{no}\dn}(t)\to0\quad\mbox{as $t\to\infty$} 
\end{equation}
is a necessary property of $p_{\mathrm{no}\dn}(t)$.
As a consequence, all eigenvalues of  $\cL_{\eta}$ must have 
a negative real part.

\begin{figure}[!t]
\centering
\includegraphics[bb=141 500 476 620]{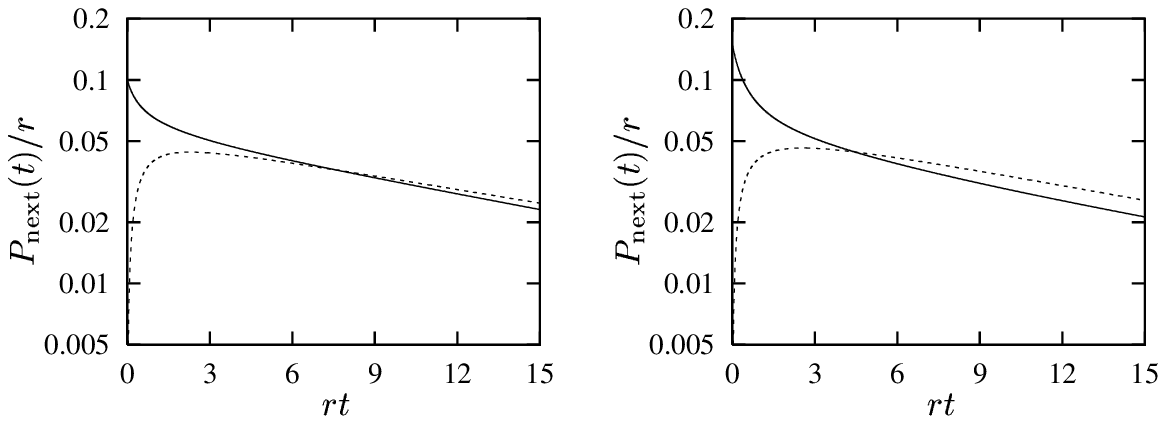}
\caption[Waiting time distributions for parity measurements]{\label{DMEfig:9}%
Waiting time distributions for parity measurements on an unpumped resonator.
The solid $\frac{\qquad}{}$ lines show $P_{\mathrm{next}}(t)/r$ for
$\DN$ clicks after $\DN$ clicks (left) and $\UP$ clicks after $\UP$ 
clicks (right) as functions of $rt$. 
The dashed $---$ lines refer to $\UP$ clicks after $\DN$ clicks (left)
and  $\DN$ clicks after $\UP$ clicks (right).
In these logarithmic plots, straight lines would correspond to 
the Poissonian statistics of uncorrelated clicks}
\end{figure}

Putting all things together we obtain
\begin{equation}
  \label{DMEeq:44h}
  P_{\mathrm{next}\dn}(t)=r\eta_{\dn}
\frac{\DMEtr{\cA\Exp{(\cL_0-r\eta_{\srdn}\cA)t}\cB\varrho^{\mathrm{(ss)}}}}
{\DMEtr{\cB\varrho^{\mathrm{(ss)}}}}
\end{equation}
for the waiting time distribution for the next $\DN$ click after a $\UP$
click.
Analogous expressions apply for the next $\UP$ click after a $\DN$ click,
the next $\DN$ click after a $\DN$ click, and so forth.
As a basic check of consistency we consider the situation in which $\cA$ and
$\cB$ are just multiples of the identity,
\begin{equation}
  \label{DMEeq:44i}
  \cA\varrho_t=q\varrho_t\;,\qquad\cB\varrho_t=(1-q)\varrho_t
\qquad\mbox{with}\quad 0<q<1\;.
\end{equation}
Then the detector clicks are not correlated at all and the waiting time
distribution \DMEeq{44h} should be Poissonian,
\index{waiting time statistics!Poissonian \DMEmain}
\index{Poissonian statistics!for waiting times}
\begin{equation}
   P_{\mathrm{next}\dn}(t)=r\eta_{\dn}q\Exp{-r\eta_{\dn}qt}\;,
\end{equation}
and this is indeed what we get from \DMEeq{44h} for \DMEeq{44i}.

Figure~\ref{DMEfig:9} shows the waiting time distributions to the parity
measurements of Figs.~\ref{DMEfig:6}--\ref{DMEfig:8}. 
Other examples are presented in some figures of \cite{DMEref:BESW94}.
\DMEendIND{waiting time statistics}

\subsection{Counting Statistics}
\DMEbegIND{counting statistics}
Yet another question is this:
What is the probability $w_n(t)$ for detecting $n$ atoms in state $\DN$ during
a period of duration $t$?
We pay no attention to $\UP$ clicks and, therefore, have $\eta_{\up}=0$ in the
nonlinear master equation \DMEeq{38c} for the evolution between $\DN$ clicks.

Probability $w_0(t)$ is the no-click probability of Sect.~\ref{DMEsec:5b}, 
\begin{equation}
  w_0=\DMEtr{\Exp{\cL_{\eta}t}\varrho^{\mathrm{(ss)}}}
\qquad\mbox{with}\qquad
\cC=\eta_{\dn}\cA\;,
\end{equation}
where $\varrho_{t=0}=\varrho^{\mathrm{(ss)}}$ in \DMEeq{44e} is appropriate
now.
The one-click probability $w_1(t)$ is given by
\begin{equation}
  \label{DMEeq:45b}
  w_1(t)=r\int_0^t\D t'\,\DMEtr{\Exp{\cL_{\eta}(t-t')}\varrho_{t=t'+0}}
\DMEtr{\cC\varrho_{t=t'-0}}
\DMEtr{\Exp{\cL_{\eta}t'}\varrho^{\mathrm{(ss)}}}\;,
\end{equation}
where the last factor is the probability $w_0(t')$ for no click before $t'$,
the first factor is the probability  for no click after $t'$, and
\begin{equation}
  r\D t'\,\DMEtr{\cC\varrho_{t=t'-0}}
\end{equation}
is the probability for a click at $t'$.
In accordance with \DMEeq{38e} and \DMEeq{39d}, the statistical operators just
before and after the click at $t'$ are
\begin{equation}
  \varrho_{t=t'-0}=\frac{\Exp{\cL_{\eta}t'}\varrho^{\mathrm{(ss)}}}
                        {\DMEtr{\Exp{\cL_{\eta}t'}\varrho^{\mathrm{(ss)}}}}
\end{equation}
and
\begin{equation}
  \varrho_{t=t'+0}=\frac{\cA\varrho_{t=t'-0}}{\DMEtr{\cA\varrho_{t=t'-0}}}
=\frac{\cC\varrho_{t=t'-0}}{\DMEtr{\cC\varrho_{t=t'-0}}}\;,
\end{equation}
respectively, so that
\begin{equation}
  \varrho_{t=t'+0}\DMEtr{\cC\varrho_{t=t'-0}}
\DMEtr{\Exp{\cL_{\eta}t'}\varrho^{\mathrm{(ss)}}}
=\cC\Exp{\cL_{\eta}t'}\varrho^{\mathrm{(ss)}}\;,
\end{equation}
and  a remarkable simplification happens, inasmuch as
\begin{equation}
  \label{DMEeq:45g}
  w_1(t)=\int_0^t\D t'\,\DMEtr{\Exp{\cL_{\eta}(t-t')}r\cC
                          \Exp{\cL_{\eta}t'}\varrho^{\mathrm{(ss)}}}
\end{equation}
involves but a single trace as the equivalent replacement of 
the product of three traces with which we started in \DMEeq{45b}.

Upon writing \DMEeq{45g} as a double integral
\begin{equation}
   w_1(t)=\int_0^{\infty}\D t_1\,\int_0^{\infty}\D t_0\,\delta(t_0+t_1-t)
\DMEtr{\Exp{\cL_{\eta}t_1}r\cC\Exp{\cL_{\eta}t_0}\varrho^{\mathrm{(ss)}}}
\end{equation}
it is reasonably obvious (and can be demonstrated by a simple induction)
that
\begin{eqnarray}
  \label{DMEeq:46b}
    w_n(t)&=&\int_0^{\infty}\!\!\D t_n\,\int_0^{\infty}\!\!\D t_{n-1}\,
           \cdots\int_0^{\infty}\!\!\D t_1\,\int_0^{\infty}\!\!\D t_0\,
           \delta(t_0+t_1+\cdots+t_{n-1}+t_n-t)
\nonumber\\
&&\mbox{\quad}\times
\DMEtr{\Exp{\cL_{\eta}t_n}r\cC\Exp{\cL_{\eta}t_{n-1}}r\cC\cdots
\Exp{\cL_{\eta}t_1}r\cC\Exp{\cL_{\eta}t_0}
\varrho^{\mathrm{(ss)}}}
\end{eqnarray}
or
\begin{equation}
  w_n(t)=\DMEtr{\cW_n(t)\varrho^{\mathrm{(ss)}}}\;.
\end{equation}
The operator $\cW_n(t)$ thus introduced,
\begin{equation}
  \cW_n(t)=\int_0^{\infty}\!\!\D t_n\,\cdots\int_0^{\infty}\!\!\D t_0\, 
\delta(t_0+\cdots+t_n-t)\Exp{\cL_{\eta}t_n}r\cC\Exp{\cL_{\eta}t_{n-1}}\cdots
r\cC\Exp{\cL_{\eta}t_0}\;,
\end{equation}
obeys the recurrence relation
\begin{equation}
  \cW_n(t)=\int_0^t\D t'\,\cW_0(t-t')\,r\cC\,\cW_{n-1}(t')
\qquad\mbox{for $n=1,2,3,\dots$}
\end{equation}
that commences with
\begin{equation}
  \cW_0(t)=\Exp{\cL_{\eta}t}\;.
\end{equation}

As always, we'll find it convenient to use a generating function,
\begin{equation}
  \cW(x,t)=\sum_{n=0}^{\infty}x^n\cW_n(t)\;.
\end{equation}
The recurrence relation for the $\cW_n(t)$'s then turns into an integral
equation for $\cW(x,t)$,
\begin{equation}
  \cW(x,t)=\cW_0(t)+\int_0^t\D t'\,\cW_0(t-t')\, xr\cC\,\cW(x,t')\;.
\end{equation}

The standard technique for handling such finite-range convolutions utilizes
Laplace transforms because the familiar identity
\begin{equation}
\int_0^{\infty}\!\!\D t\,\Exp{-\gamma t}\int_0^t\!\D t'\,f(t-t')g(t')
=\int_0^{\infty}\!\!\D t\,\Exp{-\gamma t}f(t)
\int_0^{\infty}\!\!\D t'\,\Exp{-\gamma t'}g(t')
\end{equation}
leads to a factorization.
With
\begin{equation}
  \label{DMEeq:47d}
\int_0^{\infty}\!\!\D t\,\Exp{-\gamma t}\cW_0(t)
=\bigl(\gamma-\cL_{\eta}\bigr)^{-1}  
\end{equation}
this gets us to
\begin{eqnarray}
\int_0^{\infty}\!\!\D t\,\Exp{-\gamma t}\cW(x,t)
&=&\bigl[1-\bigl(\gamma-\cL_{\eta}\bigr)^{-1}xr\cC\bigr]^{-1} 
\bigl(\gamma-\cL_{\eta}\bigr)^{-1}
\nonumber\\
&=&\Bigl(\bigl(\gamma-\cL_{\eta}\bigr)
\bigl[1-\bigl(\gamma-\cL_{\eta}\bigr)^{-1}xr\cC\bigr]\Bigr)^{-1}  
\nonumber\\
&=&\bigl(\gamma-\cL_{\eta}-xr\cC\bigr)^{-1}\;,
\end{eqnarray}
and the inverse Laplace transform is elementary,
\begin{equation}
  \cW(x,t)=\Exp{(\cL_{\eta}+xr\cC)t}\;.
\end{equation}
As we noted at \DMEeq{44g}, all eigenvalues of $\cL_{\eta}$ have negative real
parts, and so the Laplace transform \DMEeq{47d} of $\cW_0(t)$
surely exists for $\gamma\geq0$.
The same remark applies to $\cW(x,t)$ for $-(1-\eta_{\dn})/\eta_{\dn}<x<1$
because
\begin{equation}
  \cL_{\eta}+xr\cC=\cL_0-(1-x)r\cC=\cL_0-r(1-x)\eta_{\dn}\cA
\end{equation}
is $ \cL_{\eta}$ with $\eta_{\dn}$ replaced by $(1-x)\eta_{\dn}$ so that 
$\cL_{\eta}+xr\cC$ is just another operator of the same kind as $\cL_{\eta}$
if the ``effective detection efficiency'' $(1-x)\eta_{\dn}$ is in the range
$0\cdots1$, which restricts $x$ to the range stated above.

After putting things together we obtain
\begin{equation}
\label{DMEeq:48a}
\sum_{n=0}^{\infty}x^nw_n(t)
=\DMEtr{\Exp{(\cL_{\eta}+xr\cC)t}\varrho^{\mathrm{(ss)}}}
\end{equation}
as the generating function for the counting probabilities $w_n(t)$.
\index{generating function!for counting probabilities}%
\index{counting statistics!generating function}%
In view of what we observed above about $\cL_{\eta}+xr\cC$,
the right-hand side of \DMEeq{48a}
is equal to the no-click probability $w_0(t)$ for detection 
efficiency $(1-x)\eta_{\dn}$.
Accordingly, $w_0(t)$ determines all $w_n(t)$ through its dependence on
$\eta_{\dn}$.
As an immediate consequence of this observation, we get a statement about the
moments of the counting statistics,
\begin{equation}
  \sum_{n=0}^{\infty}{n \choose k}w_n(t)
=\frac{1}{k!}\left(\frac{\partial}{\partial x}\right)^k
\sum_{n=0}^{\infty}x^nw_n(t)\Big|_{x=1}
=\eta_{\dn}^k
\left[\eta_{\dn}^{-k}w_k(t)\Big|_{\eta_{\srdn}\to0}\right]\;.
\end{equation}
For $k=0$, we check the normalization,
\begin{equation}
  \sum_{n=0}^{\infty}w_n(t)=\DMEtr{\Exp{\cL_0t}\varrho^{\mathrm{(ss)}}}=1\;;
\end{equation}
for $k=1$, we get the average number of $\DN$ clicks,
\begin{eqnarray}
  \label{DMEeq:48d}
  \sum_{n=0}^{\infty}nw_n(t)&=&
\int_0^t\D t'\,\DMEtr{\Exp{\cL_0(t-t')}r\eta_{\dn}\cA
                          \Exp{\cL_0t'}\varrho^{\mathrm{(ss)}}}
\nonumber\\
&=&\eta_{\dn}rt\DMEtr{\cA\varrho^{\mathrm{(ss)}}}\;,
\end{eqnarray}
which can be understood as a statement about the ergodicity of the process
\cite{DMEref:BESW94};
and for $k=2$, we learn something about the variance of the counting
statistics,
\begin{eqnarray}
  \label{DMEeq:48e}
&&  \sum_{n=0}^{\infty}n(n-1)w_n(t)
\nonumber\\
&&\qquad=2
\int_0^t\D t'\,\int_0^{t'}\D t''\,
\DMEtr{\Exp{\cL_0(t-t')}r\eta_{\dn}\cA\Exp{\cL_0(t'-t'')}r\eta_{\dn}\cA
                          \Exp{\cL_0t''}\varrho^{\mathrm{(ss)}}}
\nonumber\\[1ex]
&&\qquad=
(\eta_{\dn}rt)^2\DMEtr{\cA E(\cL_0t)\cA\varrho^{\mathrm{(ss)}}}\;,
\end{eqnarray}
with
\begin{equation}
  E(y)=\frac{2}{y^2}\bigl(\Exp{y}-1-y\bigr)\;.
\end{equation}
In traces of integrals such as \DMEeq{48d} and \DMEeq{48e}, the exponential on
the left and on the right can be ignored because $\cL_0$ is trace conserving
and $\varrho^{\mathrm{(ss)}}$ is its right eigenvector to eigenvalue zero. 

Here, too, we get Poissonian statistics for \DMEeq{44i}, namely
\index{counting statistics!Poissonian \DMEmain}
\index{Poissonian statistics!for counting probabilities}
\begin{equation}
  \label{DMEeq:49a}
  \sum_{n=0}^{\infty}x^nw_n(t)=\Exp{-(1-x)r\eta_{\srdn}qt}\;,\qquad
w_n(t)=\frac{(r\eta_{\dn}qt)^n}{n!}\Exp{-r\eta_{\srdn}qt}\;,
\end{equation}
for which
\begin{equation}
  \sum_{n=0}^{\infty}{n \choose k}w_n(t)=\frac{(r\eta_{\dn}qt)^k}{k!}\;.
\end{equation}
In particular, we note that
\begin{equation}  
\sum_{n=0}^{\infty}n(n-1)w_n(t)
=\left[\sum_{n=0}^{\infty}nw_n(t)\right]^2
  \label{DMEeq:49c}
\end{equation}
holds for the Poissonian counting statistics \DMEeq{49a}.

A convenient, yet rough, measure for the deviation from Poissonian statistics
is the so-called \DMEind{Fano--Mandel factor} $Q(t)$,
\begin{equation}
  Q=\frac{\displaystyle\sum_{n=0}^{\infty}n(n-1)w_n}
            {\displaystyle\sum_{n=0}^{\infty}nw_n}
      -\sum_{n=0}^{\infty}nw_n\;,
\end{equation}
a normalized variance. 
The normalization is such that $Q=0$ for Poissonian counting statistics, as
one verifies easily with \DMEeq{49c}.
For ${-1\leq Q<0}$ one speaks of sub-Poissonian statistics, and of
super-Poissonian statistics for $Q>0$.

\begin{figure}[!t]
\centering
\includegraphics[bb=160 420 455 630]{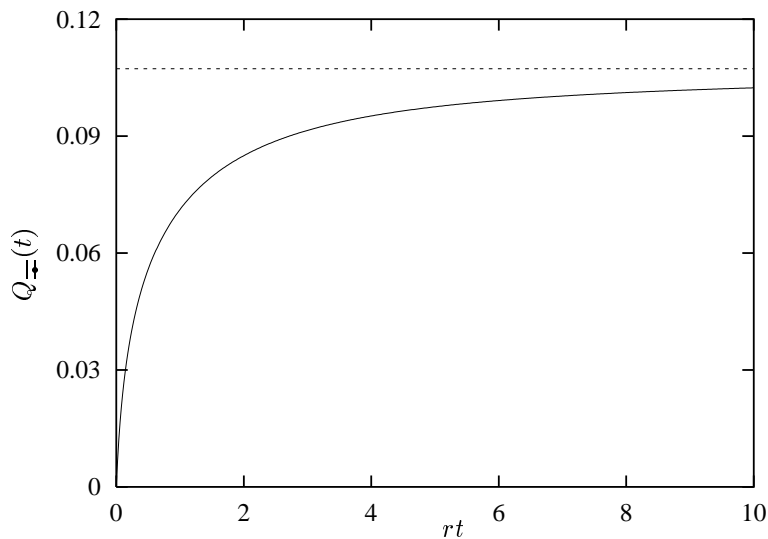}
\caption[Fano--Mandel factor for parity measurements]{\label{DMEfig:10}%
Fano--Mandel factor for the counting statistics in parity measurements 
on an unpumped resonator.
The solid $\frac{\qquad}{}$ line shows $Q_{\dn}(t)$ of \DMEeq{50c},
the horizontal dashed $---$ line is the asymptotic value of \DMEeq{50d}}
\end{figure}

For the count of $\DN$ clicks, \DMEeq{48d} and \DMEeq{48e} give
\begin{equation}
  Q_{\dn}(t)=\eta_{\dn}rt\left[
\frac{\DMEtr{\cA E(\cL_0t)\cA\varrho^{\mathrm{(ss)}}}}
     {\DMEtr{\cA\varrho^{\mathrm{(ss)}}}}
-\DMEtr{\cA\varrho^{\mathrm{(ss)}}}\right]\;.
\end{equation}
We use the damping bases to get a tractable numerical expression,
\begin{equation}
  \label{DMEeq:50b}
  Q_{\dn}(t)=\eta_{\dn}rt\sum_{n=0}^{\infty}\sum_{k=-\infty}^{\infty}
\bigl(1-\delta_{n,0}\delta_{k,0}\bigr)
\frac{\DMEtr{\cA\varrho_n^{(k)}}E(\lambda_n^{(k)}t)
\DMEtr{\check{\varrho}_n^{(k)}\cA\varrho_0^{(0)}}}
{\DMEtr{\cA\varrho_0^{(0)}}}\;,
\end{equation}
where the $n=0,k=0$ term is removed since
$\varrho_0^{(0)}=\varrho^{\mathrm{(ss)}}$, $\check{\varrho}_0^{(0)}=1$,
$\lambda_0^{(0)}=0$, and $E(y=0)=1$.
For the parity measurements of Figs.~\ref{DMEfig:6}--\ref{DMEfig:9}, one can
evaluate the sum and gets
\begin{equation}
  \label{DMEeq:50c}
 Q_{\dn}(t)=\frac{\eta_{\dn}r}{(\nu+1)(2\nu+1)}
\int_0^t\frac{\D t'}{2At}
\ln\Bigl(1+4\nu(\nu+1)\bigl(1-\Exp{-At'}\bigr)\Bigr)\;,  
\end{equation}
which is plotted in Fig.~\ref{DMEfig:10}.
For $At\ll1$ and $At\gg1$ the limiting forms
\begin{equation}
  \label{DMEeq:50d}
   Q_{\dn}(t)\simeq\left\{
     \begin{array}{l@{\mbox{\ for\ }}l}
\displaystyle\frac{\nu}{2\nu+1}\eta_{\dn}rt & At\ll1\;,\\[2ex]
\displaystyle\frac{\ln(2\nu+1)}{(\nu+1)(2\nu+1)}\eta_{\dn}r/A
& At\gg1\;,
     \end{array}\right.
\end{equation}
obtain, as is confirmed by Fig.~\ref{DMEfig:10}.
Other examples of Fano--Mandel factors for counting statistics are presented
in some figures of \cite{DMEref:BESW94}.

The pioneering measurement in 1990 of atom statistics in a real-life micromaser
experiment\index{micromaser!measured atom statistics}
is reported in \cite{DMEref:R+SK+W90} and linked to the photon
counting statistics in \cite{DMEref:R+W90}.  
Measured Fano--Mandel factors\index{micromaser!measured Fano--Mandel factors}
 from some later experiments can be found in
\cite{DMEref:ELBVWW98}. 
\DMEendIND{counting statistics}

\subsection*{Homework Assignments}
\addcontentsline{toc}{subsection}{\protect\numberline{}Homework Assignments}
\begin{enumerate}
\renewcommand{\labelenumi}{\textbf{\theenumi}}
\addtocounter{enumi}{\value{DMEhw}}
\item \label{DMEhw:5a}
What is the correlation function for clicks of either kind, that is: without
caring if it's a $\DN$ click or a $\UP$ click?
\item 
The two cross-correlation functions are identical 
in the example of \DMEeq{42a}--\DMEeq{42g}, see \DMEeq{42e}.
Is this always the case?  
\item 
Since the next click is bound to come sooner or later, consistency requires
that $P_{\mathrm{next}}(t)$ is normalized to unit integral.
Show that this is indeed so for  $P_{\mathrm{next}}(t)$ of \DMEeq{44f}.
\item 
A $\DN$ click happens at $t=0$.
What is the probability that the next click is a $\UP$ click?
\item 
Use the methods of Sect.~\ref{DMEsec:5b} to find an expression 
for the average waiting time between successive $\DN$ clicks, between
successive $\UP$ clicks.
\item 
Determine the short-time behavior of the various $P_{\mathrm{next}}$'s of
Fig.~\ref{DMEfig:9} and compare with the plots.
\item 
Show that the exponential function $\Exp{F}$ of an operator $F$ responds to
variations $\delta F$ in accordance with
\begin{equation}
  \label{DMEeq:51a}
  \delta\Exp{F}=\int_0^1\!\D\tau\,\Exp{\tau F}\delta F  \Exp{(1-\tau)F}\;,
\end{equation}
which epitomizes all of perturbation theory.
\item 
Use \DMEeq{51a} to extract $w_1(t)$ and $w_2(t)$ from \DMEeq{48a}.
Compare with \DMEeq{46b}.
\item 
Consider the probability $w_{nm}(t)$ of detecting $n$ atoms in state $\DN$ and
$m$ atoms in state $\UP$ during a period of duration $t$.
Show that\index{generating function!for counting probabilities}
\index{counting statistics!generating function}
\begin{equation}
  \sum_{n,m=0}^{\infty}x^ny^mw_{nm}(t)
=\DMEtr{\Exp{(\cL_{\eta}+xr\eta_{\srdn}\cA+yr\eta_{\srup}\cB)t}
\varrho^{\mathrm{(ss)}}}
\end{equation}
is the appropriate generalization of the generating function \DMEeq{48a}.
\item 
For $\cA$ of \DMEeq{39a}, evaluate the $k=0$ traces in \DMEeq{50b} and so
confirm \DMEeq{50c}.
\item 
Find the leading correction to the approximation given in \DMEeq{50d} 
for $At\gg1$.
\end{enumerate}



\section*{Acknowledgments}
\addcontentsline{toc}{section}{\protect\numberline{}Acknowledgments}
We congratulate the organizers of the CohEvol seminar and workshop in Dresden.
They succeeded in putting together a stimulating meeting in a most enjoyable
atmosphere. 

Much of the work reported in these lectures was done at the 
Max-Planck-Institut f\"ur Quantenoptik in Garching.
BGE is greatly indebted to Herbert Walther for his generous hospitality and
support over all those years.

The evolution of the damping-basis method from a crude idea to a powerful tool
would not have occurred without the ingenuity and stamina of Hans Briegel.
BGE will always remember this collaboration with great affection.
 
At the time of the CohEvol seminar and workshop,
BGE enjoyed the hospitality of the Atominstitut in Vienna.
He wishes to thank Helmut Rauch and Gerald Badurek for the splendid
environment they provided, and the Technical University of Vienna for
financial support.
BGE is equally grateful to Goong Chen and Marlan Scully for the visiting
professorship they arranged at Texas A\&M University, where these 
notes were finalized.

GM wishes to express her sincere gratitude for the strong encouragement and
support by Herbert Walther and the members of his Garching group.
And she thanks all participants of the workshop for the many memorable hours
spent together.

\section*{Appendix}
\addcontentsline{toc}{section}{\protect\numberline{}Appendix}
Here are some facts about special functions that are useful for homework
assignments \ref{DMEhw:2b} and \ref{DMEhw:2e}. 
The expansion of \DMEeq{20a} in powers of 
$\Exp{-\I\omega t}\propto\bigl[\alpha(t)/\alpha^*(t)\bigr]\power{\thalf}$ 
is done with the aid of
\begin{equation}
  \Exp{\thalf x(y-1/y)}=\sum_{k=-\infty}^{\infty}y^k\mathrm{J}_k(x)\;,
\end{equation}
the most important generating function for Bessel functions of integer order,
\index{generating function!for Bessel functions}
\begin{equation}
  \mathrm{J}_k(x)=(-1)^k\mathrm{J}_{-k}(x)=\I^{|k|-k}\mathrm{J}_{|k|}(x)
=\I^{|k|-k}\sum_{m=0}^{\infty}\frac{(-1)^m}{m!\,(m+|k|)!}
\left(\thalf x\right)^{|k|+2m}\;.
\end{equation}
They in turn act as a generating function for Laguerre polynomials,
\index{generating function!for Laguerre polynomials}
\begin{equation}
\mathrm{J}_{|k|}\bigl(2\sqrt{xy}\,\bigr) =(xy)\power{\thalf|k|}\Exp{-y}
\sum_{n=0}^{\infty}\frac{y^n}{(n+|k|)!}\mathrm{L}_n^{(|k|)}(x)\;. 
\end{equation}
After a suitable Laplace transform this becomes
\begin{equation}
  (1+y)^{-|k|-1}\Exp{\frac{xy}{1+y}}=\sum_{n=0}^{\infty}(-y)^n
\mathrm{L}_n^{(|k|)}(x)\;,
\end{equation}
which is another useful generating function for the Laguerre polynomials
\index{generating function!for Laguerre polynomials}
\begin{equation}
  \mathrm{L}_n^{(|k|)}(x)
=\sum_{m=0}^n{{n+|k|}\choose{m+|k|}}\frac{(-x)^m}{m!}\;.
\end{equation}

The integral relations
\begin{equation}
  \mathrm{L}_n^{(|k|)}(x)=\frac{1}{n!}\Exp{x}x\power{-\thalf|k|}
\int_0^{\infty}\!\D y\,\Exp{-y}y\power{n+\thalf|k|} 
\mathrm{J}_{|k|}\bigl(2\sqrt{xy}\,\bigr)
\end{equation}
and
\begin{equation}
\int_0^{\infty}\!\D y\,\Exp{-y}
\mathrm{J}_{|k|}\bigl(2\sqrt{uy}\,\bigr)
\mathrm{J}_{|k|}\bigl(2\sqrt{vy}\,\bigr)=\Exp{-(u+v)}
\mathrm{I}_{k}\bigl(2\sqrt{uv}\,\bigr)  
\end{equation}
are worth knowing, where
\begin{equation}
 \mathrm{I}_{k}(x)=\mathrm{I}_{-k}(x)=\I^{-k}\mathrm{J}_{k}(\I x)
\end{equation}
are modified Bessel functions of integer order.
As a preparation for homework assignment \ref{DMEhw:2b} you might want to
derive first
\begin{equation}
  \sum_{n=0}^{\infty}\frac{n!}{(n+|k|)!}x^n
  \mathrm{L}_n^{(|k|)}(y)  \mathrm{L}_n^{(|k|)}(z)
=\frac{(xyz)\power{-\thalf|k|}}{1-x}\Exp{-\frac{x}{1-x}(y+z)}
 \mathrm{I}_{k}\left(2\frac{\sqrt{xyz}}{1-x}\right)
\end{equation}
by combining some of these relations fittingly.


\clearpage
\flushbottom
\printindex

\end{document}